\documentclass[aip,jmp,gsraphicx,nofootinbib]{revtex4-2}
\draft 
\usepackage{amssymb}
\usepackage[colorlinks]{hyperref}
\usepackage{float} 
\usepackage{marginnote}
\usepackage{mathtools}	
\usepackage{xcolor}

\begin{document}
\title{Spacetime triple wormhole} 
\author{Vincent Herr}
\email[]{vincent.herr@ucdenver.edu}
\author{Aim\'e Fournier}
\affiliation{University of Colorado Denver, Department of Mathematical and Statistical Sciences}
\author{Andrew J.S. Hamilton}
\affiliation{University of Colorado Boulder, JILA, and Department of Astrophysical and Planetary Sciences}

\date{5/22/2026}

\begin{abstract}
\textbf{Abstract}
We describe a multi-neck spacetime wormhole with a simple metric tensor and a simple injective map without coordinate patching. An intra-universe, non-thin-shell, non-spherically-symmetric 3-neck spacetime wormhole is geometrically constructed by spherically inverting a 3-torus. We place the resulting Dupin hypercyclide in a synchronous reference frame.
The three necks are arranged around a central point and satisfy
topological and geometric spacetime wormhole definitions. 
Asserting this metric tensor as an exact solution of Einstein’s field equations in global coordinates generates  diagonal Ricci and stress-energy tensors, and a Riemann curvature tensor with only six nonzero entries.
The local inertial frame at every point of the coordinate system is comoving with the triple wormhole.
This non-vacuum solution answers affirmatively the question posed by Einstein and Rosen (1935) of whether  or not multi-neck solutions exist.
The wormhole solution contains negative energy density as is expected to hold the necks open; however, geodesic paths through each neck exist which encounter only positive energy density. 
The spatial manifold is a trivariate Dupin hypercyclide.
The spherically inverted equal-radii 3-torus is unbounded, asymptotically flat, and admits a global isothermal coordinate system that further simplifies the curvature tensors.
\end{abstract}


\maketitle

\hypersetup{linkcolor=blue} 


\section{Introduction}\label{s:intro}

Wormhole solutions to Einstein's field equations (EFE, see \S \ref{s:APXefe}) have been studied for many decades. In 1935 Einstein and Rosen\cite{ER} modified the mathematical solution for a black hole to join two such solutions by a ``bridge,'' proposing this as a geometrical particle model. While anticipating models with multiple bridges, they stated,
``For the present one cannot even know whether regular solutions with more than one bridge exist at all.'' Since then, multi-neck wormhole models have been geometrically constructed using multiple solutions joined by coordinate patching\cite{misner-1963}. The present work offers a global metric solution for a three-neck spacetime wormhole without coordinate patching.

\citet{misner-wheeler-1957}[p. 532] first coined the term ``wormhole'' in 1957 to describe a doubly-connected topology of spacetime. Misner\cite{misner-1960} in 1960 mapped a single-neck spacetime wormhole by placing an asymptotically flat transformation of the ``3-dimensional doughnut'' $S^1\times S^2$, placing the ``metric of flat space in bispherical coordinates'' in a \textit{synchronous-frame} spacetime (see \citet[section 99]{landau-lifshitz-1971}).
Kanumilli\cite{kanumilli-pope-2018} and Pope generalized this in 2018 to model single-neck wormholes based on $S^1\times S^{n-1}$ in $n+1$-dimensional spacetimes. Our present work models a three-neck wormhole in 3+1-dimensional spacetime using Pinkall's\cite{pinkall-1985B} geometric framework for Dupin hypercyclides. Misner's\citep[1960]{misner-1960} wormhole geometry and that of our paper are both available via  construction (2) in Section 7 of Pinkall\cite{pinkall-1985B}.

Our model is size-scalable by the factor $\sigma$ in the mapping equations of Section \ref{s:proofs}. In 1995 \citet{visser-1995}[p. xix] wrote, ``There are plausible physical arguments that Lorentzian 
(locally Minkowski) 
wormholes should exist at least at a microscopic scale (sizes of order the Planck length, about $10^{-35}$ meters).'' 
Interest in spacetime wormholes increased after 2013 when \citet{maldacena-susskind-2013} proposed a micro-wormhole connecting two particles as a dual representation of quantum entanglement.

The purpose of this paper is to expose a geometric model of multiply-connected (multi-neck) non-spherically symmetric spacetime wormholes formed by appending time to Dupin hypercyclides of a different topology than Misner\cite{misner-1960} analyzed in 1960, specifically by inverting $n$-tori $(S^1\times S^1\times\cdots\times S^1)$ embedded for mathematical convenience in higher-dimensional Euclidean space. We examine the characteristics of this model in light of its stress-energy and other tensors. We do not address any questions of interpreting wormholes as particles. 

The geometry of Dupin hypercyclides and the topology of spacetime wormholes are discussed in \S\ref{s:bg}. 
The novel parts of the present work detailed in \S\ref{s:proofs} include
1) a non-spherically symmetric multi-neck spacetime wormhole in standard unmodified GR with global metric and no coordinate patching,
2) proof that it satisfies topological and geometric definitions of spacetime wormhole,
3) proof that it generates a diagonal stress-energy tensor, i.e., a physical scenario with zero spatial momentum and zero shear stresses,
4) asymptotically Minkowski triple wormhole example, 
5) no singularity,
and 6) a triple wormhole example with globally isothermal coordinate system.
Generally, isothermal coordinates on a Riemannian manifold are coordinates whose metric is locally conformal to the Euclidean metric, 
$ds^2=\phi\left(dx_1^2+\cdots+dx_n^2\right)$, where 
$\phi({\bf x})>0$ 
is a smooth function. A globally isothermal metric uses the same function $\phi$ everywhere.

For simplicity, using geometrized units (\citet[p. 36]{mtw}) in which the speed of light $c=1$ and Newton's gravitational constant $G=1$, the EFE (\ref{e:efe}) are a set of ten nonlinear partial differential equations (PDEs):
\begin{equation}\label{e:efe}
    R_{\mu\nu}-\frac{1}{2}g_{\mu\nu}R
    =8\pi 
    T_{\mu\nu} 
\end{equation}
where $R_{\mu\nu}$ is the Ricci tensor, $R$ its trace,
and $g_{\mu\nu}$ the metric tensor.
The left-hand side describes the compressive part of curvature of spacetime at each point. In the right-hand side, the stress-energy tensor $T_{\mu\nu}$ describes material conditions at each point: energy density, momentum, and normal and shear stresses.
We use the Einstein summation convention except where noted.

Given some physical scenario described in the EFE right-hand side, solving the PDEs in the usual sense would mean finding the metric tensor $g_{\mu\nu}$. 
Each component $R_{\kappa\lambda}$ of the Ricci curvature tensor on the left-hand side is a nonlinear function of all the $g_{\mu\nu}$ and their 1st derivatives and is linear in their 2nd derivatives. 
There are only a small number of cases where Einstein's equations have been solved\cite{stephani-etal-2003} in this sense, right to left, finding the $g_{\mu\nu}$ from a specified $T_{\mu\nu}$.
If no physical restrictions are placed on the stress-energy tensor, then any proposed metric might be called an ``exact solution,'' but in general, a given metric will describe a non-physical scenario of no interest (\citet[p. 3]{Choquet-Bruhat-2003}). 
The present work follows the more common path (left to right) to an exact solution, first formulating a metric by which the field equations describe a physical scenario of particular interest. 

Non-vacuum solutions describing certain potentially physical scenarios such as spacetime wormholes are controversial, having not yet been observed in nature. One objection to their possible existence is that, in the unmodified EFE, spacetime wormhole metrics (including ours) imply unknown forms of exotic matter, e.g., including negative energy density. The Casimir force\cite{casimir-1948} has been experimentally observed and measured at a very small scale, but as of this writing, its interpretation as evidence of negative density also remains controversial\cite{lamoreaux-2005}.

The 3-torus can be embedded and inverted in 
$\mathbb{R}^4$ by a standard algebraic implicit formula of the form 
$\tau(x^1,\cdots,x^4)=0$, with 
$3\times3$ metric tensor 
$(g_{\alpha\beta})$ given as usual by the inner products of three orthogonal tangent 4-vectors in 
$\text{null}\nabla\tau$.
Conformal 4-D inversion 
$(x^1,\cdots,x^4)\mapsto(y^1,\cdots,y^4)$ 
induces a 3-cyclide metric 
$\lambda^2 g_{\alpha\beta}$ with conformal factor 
$\lambda$ derived as usual from the determinant of the $\partial y^\alpha/\partial x^\beta$.
Two of us (AH and AF) derived that 
$\lambda^2 g_{\alpha\beta}$ leads to a diagonal Einstein tensor. The local inertial frame is comoving with the triple wormhole.
The special case of inverting a flat 3-torus isometrically embedded in
$\mathbb{R}^6$ resulted in the simpler tensor expressions of this paper. 

We do not anticipate finding a stationary triple wormhole in nature with the metric of this paper. Our goal is to expose a multi-neck spacetime wormhole model with especially simple tensor formulations as a starting place for developing more compelling variations, and also as a potentially useful tool for teaching differential geometry and general relativity.  

\section{Background}\label{s:bg}

\subsection{Dupin cyclides}\label{s:dupin}

\begin{figure}
   \begin{center}
      \includegraphics[width=\linewidth]{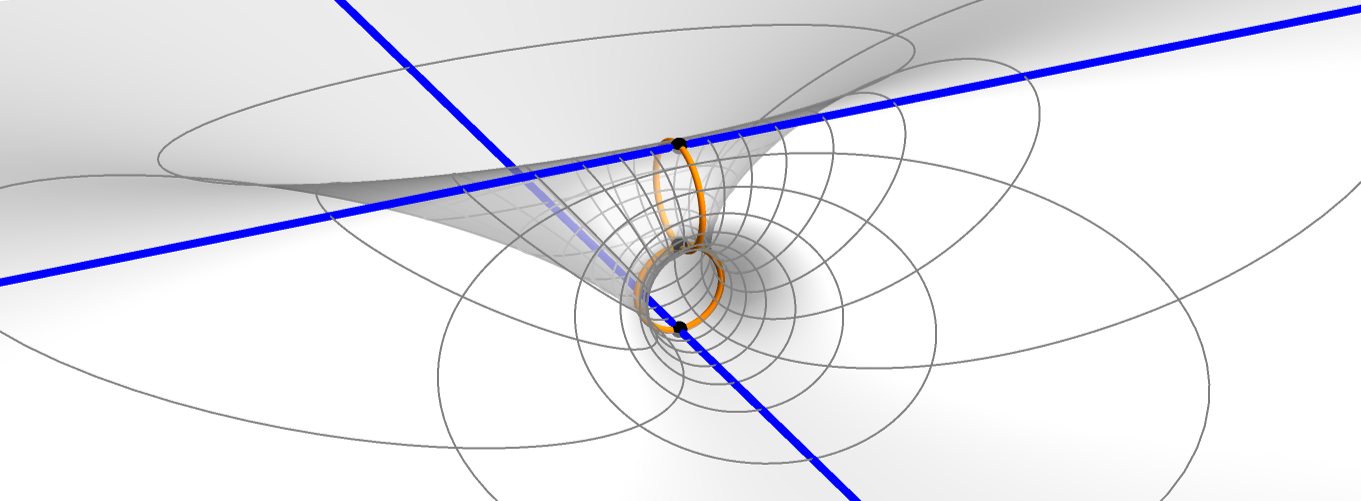}
   \end{center}
   \caption{\footnotesize The parabolic Dupin cyclide (PDC) $\mathbf{w}(\zeta^1,\zeta^2,\pi)$ is an inverted 2-torus. In terms of mapping (\ref{e:w}) the orange coordinate circles are $\mathbf{w}(\zeta^1,0,\pi)$ and $\mathbf{w}(0,\zeta^2,\pi)$ intersecting at the origin. The blue coordinate lines are $\mathbf{w}(\zeta^1,\pi,\pi)$ and $\mathbf{w}(\pi,\zeta^2,\pi)$. Other coordinate circles are gray.
   }
   \label{f:dupin}
\end{figure}
A Dupin cyclide is a canal surface covered by two families of orthogonally intersecting circles, first described\cite[1822]{dupin-1822} by Charles Dupin. 
In terms of curvature, Dupin cyclides are the simplest class of surfaces in differential geometry\cite{langevin-etal-2015} after planes and spheres.
\citet[(1909)]{eisenhart-1909} 
showed that the parabolic Dupin cyclide (PDC, Fig. \ref{f:dupin}) 
is constructed by inverting a 2-torus with one of its points being the inversion center. 

\label{s:hyperdupin}
A Dupin hypercyclide is an $n$-manifold covered by
$n$\footnotesize$>$\normalsize2 
families of orthogonal circles, which serve as convenient coordinate lines, and can be formed by spherical inversion of an $n-$torus.
\citet[(1985)]{pinkall-1985B} provided four inductive algorithms to construct Dupin cyclide hypersurfaces in $\mathbb{R}^{n+1}$ given one in $\mathbb{R}^n$. 
\citet[(1997)]{schmidt-1997} and \citet[(2020)]{cecil-chern-2020} constructed Dupin hypersurfaces with 
space-like ($ds^2>0$) and time-like ($ds^2<0$)
dimensions, but without connecting this with any physical or relativistic concepts.
This paper employs Dupin hypercyclides as models for multi-neck spacetime wormholes. Coordinate lines of constant-curvature greatly simplify their curvature tensors.

A special case of Dupin hypercyclide,
the $n$-torus is the Cartesian product of $n$ topological circles (${\bf S}^1\times\cdots\times{\bf S}^1$) forming a 
compact non-simply-connected manifold with no boundary,
embedded in at least $n+1$-dimensional Euclidean space in which each of the manifold's coordinate lines is a geometric circle.

\subsection{Flat 2-torus}
In 1873 \citet[(p. 390)]{clifford-1873} described the \emph{flat torus}: ``The geometry of this surface is the same as that of a finite parallelogram whose opposite sides are regarded as identical.''
In 1885 \citet[(p.241)]{killing-1885} mapped a flat 2-torus into $\mathbb{R}^4$, 
\begin{equation}\label{e:killing}
    {\bf x}=\left(
    a\sin\dfrac{u}{a},
    a\cos\dfrac{u}{a},
    b\sin\dfrac{v}{b},
    b\cos\dfrac{v}{b}
    \right)
\end{equation}
normalized by the denominators to give it the Euclidean line element squared
$ds^2=du^2+dv^2$.

\subsection{Topological wormhole definition}
\label{s:Def1}
The following general definition of spacetime wormhole is adapted from \citet[p.\ 90]{visser-1995}.

\label{def1}\textbf{Definition 1:  }
\emph{
In an $(n+1)$-dimensional Lorentzian spacetime, a region $\Omega$ with  topology of the form}
$\Omega\simeq\mathbb{T}\times\Sigma$
\emph{where $\Sigma$ is a compact space-like
\textbf{non-simply-connected}
$n-$manifold, with boundary having
\textbf{simply-connected}
topology of the form}
$\partial\Sigma\simeq\mathbb{S}^{n-1}$
\emph{(where the relation
$\simeq$ indicates a diffeomorphism),
and where} 
$\mathbb{T}\simeq\mathbb{R}$
\emph{represents a time-like dimension, is said to contain an
\textbf{intra-universe Lorentzian wormhole}.
A \textbf{neck} of this wormhole is a $(n+1)$-dimensional spacetime volume
$T\simeq\mathbb{T}\times\cup_{x\in[A,B]}\mathbb{M}(x)\subset\Omega$
where the $\mathbb{M}(x)$ are compact $(n-1)$-manifolds 
and where real interval $[A,B]\subset\mathbb{R}$
labels a smoothly connected continuum of compact $(n-1)$-manifolds (or sections) $\mathbb{M}(x)$ with $x\in[A,B]$, and with
$\mathbb{M}(x)$ and $\mathbb{M}(x')$ disjoint iff
$x\ne x'$.
Neck $T$ is connected (topologically glued) to $\Omega\setminus T$ only at the two disjoint 
\textbf{neck mouths}, $\mathbb{M}(A)$ and $\mathbb{M}(B)$. The $\mathbb{M}(x)$ may be simply- or non-simply connected, but must all have the same topology. This definition of neck applies only to static wormholes and only with the same timelike dimension inside the neck as outside.
}

\begin{figure}
   \begin{center}
      \includegraphics[width=1.0\linewidth]{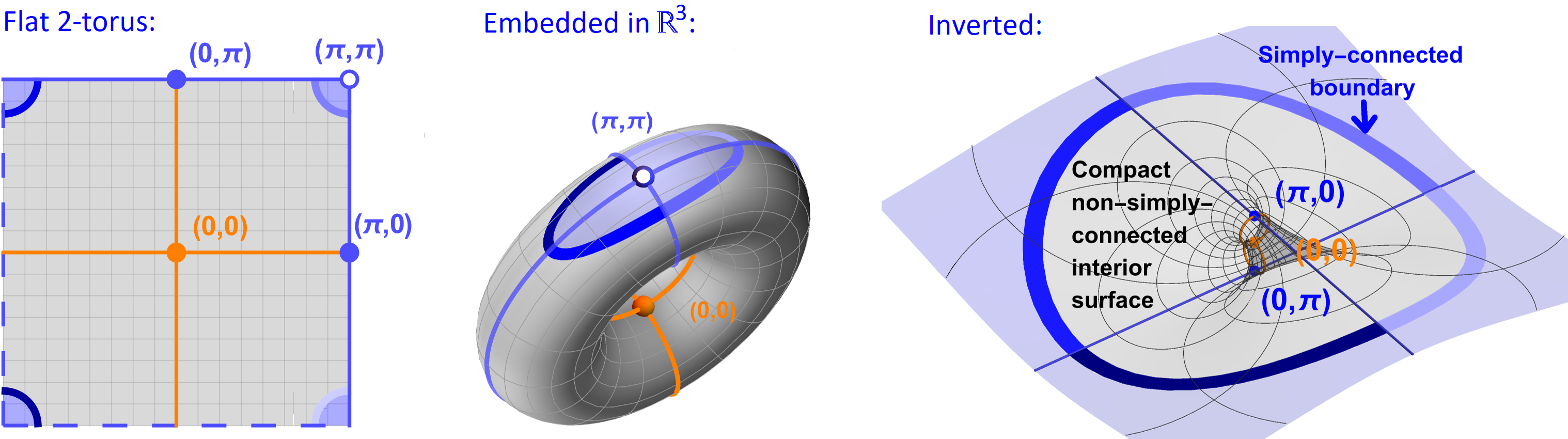}
   \end{center}
   \caption{\footnotesize Wormhole properties of a parabolic Dupin cyclide (PDC), an inverted 2-torus.
   \textbf{Left:} The flat 2-torus, represented by a square. 
   \textbf{Center:} 2-torus embedded in Euclidean space, distorts lengths, but preserves the topology of the flat 2-torus. 
   \textbf{Right:} The PDC is a spherical inversion of the 2-torus, using $(\pi,\pi)$ as the inversion point. 
   }
   \label{f:inv2Fboundary}
\end{figure}
Fig. \ref{f:inv2Fboundary} shows steps in constructing a PDC, an inverted 2-torus that meets the spatial requirements of Definition 1. The orange coordinate circles cannot be continuously shrunk to a point; the gray area is non-simply connected.
Identifying corresponding points of parallel edges of square flat torus model (left) 1) shows that the coordinate lines, while straight, nevertheless form closed loops topologically, becoming geometric circles when the torus is embedded (center) in 3-dimensional or larger Euclidean space, 
2) visually joins the four simply-connected blue quarter-discs enclosing the point $(\pi,\pi)$, and 
3) makes clear that the square model's four corners represent a single point.
After spherical inversion (right)) the nonsimply-connected gray area is enclosed by the simply-connected boundary curve (four shades of blue), illustrating the topological properties of Definition 1.

\begin{figure}
   \begin{center}
      \includegraphics[width=1.0\linewidth]{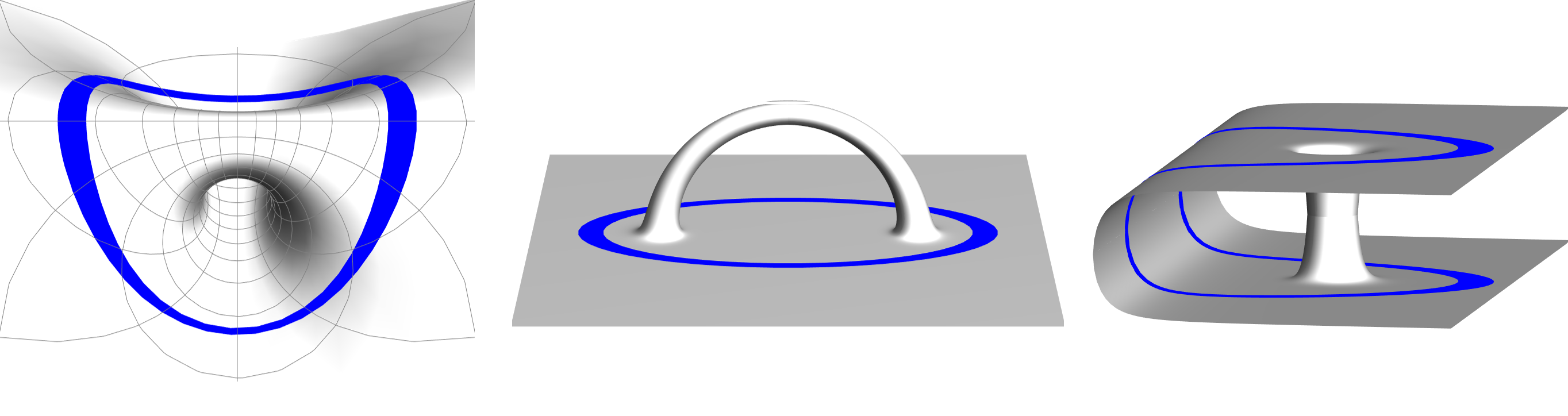}
   \end{center}
   \caption{\footnotesize
    Topological wormhole definition in two spatial dimensions: A simply-connected boundary (blue loop) encloses the doubly-connected topology of the parabolic Dupin cyclide (left), as in the the other two bridged-surface wormhole illustrations (center and right), typically omitting one spatial dimension. 
   }
   \label{f:2Dwormholes}
\end{figure}

In $n=2$ spatial dimensions we see an object with the required spatial properties, \emph{i.e.}, a closed simply-connected boundary enclosing a nonsimply-connected interior. 
Similarly, adding a unit lapse time dimension to an inverted  $n$-torus yields a $n$-neck wormhole in a $n$+1-spacetime. 
Fig. \ref{f:2Dwormholes} compares the PDC  topology with other wormhole illustrations having the same spatial properties.
Satisfying the topological definition requires global information about the connectedness of the space. In Section \ref{s:proof-topological} we prove that a triple wormhole satisfies Definition 1.

\subsection{Geometric wormhole definition}
\label{s:visserDef2}
In addition to the topological wormhole definition, Hochberg and Visser\cite{hochberg-visser-1997} generalized Morris and Thorne's\cite{morris-thorne-1988} wormhole throat definition and neck ``flare-out'' condition and offered a geometric wormhole definition for strictly local analysis. This definition applied only to static spacetimes, including nonspherically symmetric and nonasymptotically flat spacetimes, and where not all global characteristics are known. They geometrically defined a wormhole throat as a ``two-dimensional hypersurface of minimal area taken in one of the constant-time spatial slices'' using Gaussian normal coordinates, \textit{i.e.,} the throat is defined as a section through the narrowest part of a wormhole neck. The flare-out condition requires the areas of the neck sections to grow larger in both directions away from the throat. In Section \ref{s:geometric-def} we prove that a triple wormhole satisfies this geometric wormhole definition and the flare-out condition.

\section{Triple wormhole}
\label{s:proofs}
\subsection{Satisfying the topological definition of spacetime wormhole}
\label{s:proof-topological}

\label{s:f3}
\begin{figure}
   \begin{center}
      \includegraphics[width=1.0\linewidth]{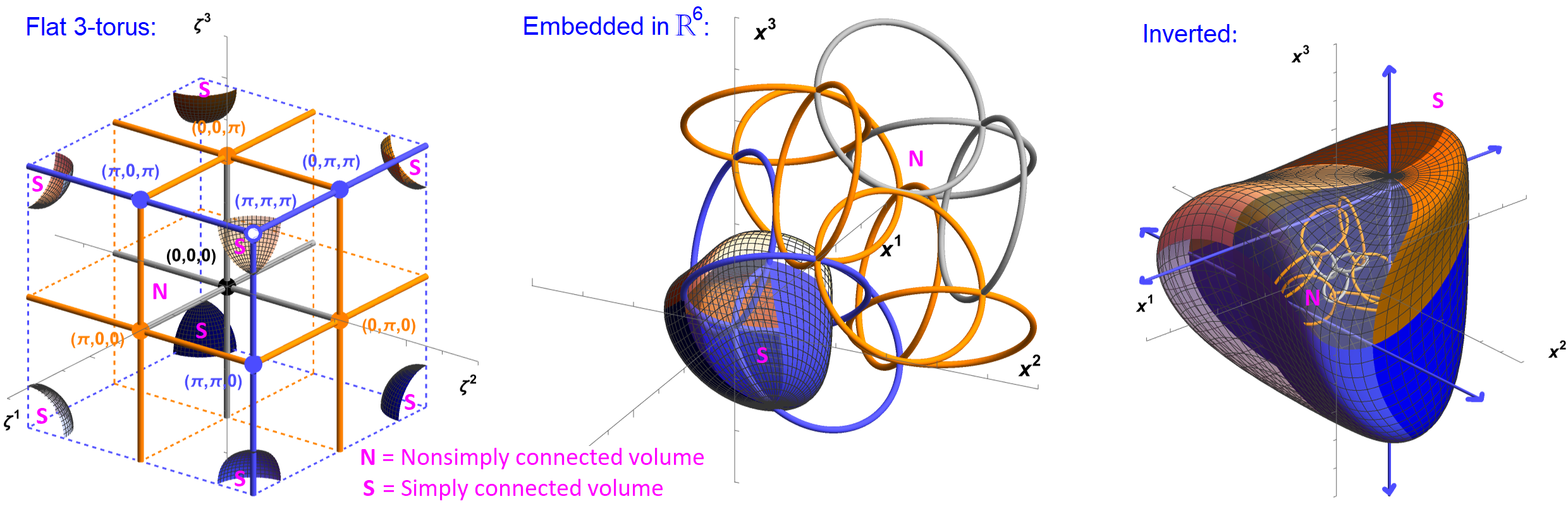}
   \end{center}
   \caption{\footnotesize 
   Wormhole properties of the \emph{inverted} flat 3-torus. 
   Same-colored wire-frame coordinate lines of the transformed 3-manifold correspond in all three images. 
   The 8-colored curved surface (partially transparent) separates each space into a simply-connected and a non-simply-connected volume (magenta ``S'' and ``N'').
   For visualization, center and right images were surjected by matrix $\mathsf{P}$ (\ref{e:P}) from $\mathbb{R}^6$ into $\mathbb{R}^3$.
   \textbf{Left:} The cube model of a 3-torus, with selected coordinate lines and their labeled intersection points.
   \textbf{Center:} Coordinate lines of the flat 3-torus spatial manifold embedded in $\mathbb{R}^6$ by mapping (\ref{e:fF3}).
   \textbf{Right:} 
   Coordinate lines of the embedded flat 3-torus inverted by mapping (\ref{e:w}).
   }
   \label{f:Ffz}
\end{figure}

We will show that the inverted image of a 3-torus embedded in spacetime is a Lorentzian
wormhole. 
First, parameterize a 3-torus with periodic coordinates $\zeta^\alpha$. For each $\zeta^\alpha$, let $\sim$ be an equivalence relation that identifies $-\pi$ with $\pi$ as on a unit circle. Then the quotient space
\begin{equation}\label{e:f3}
  \mathbb{F}^3\coloneqq
  \lbrace 
    (\zeta^1,\zeta^2,\zeta^3):\;
    -\pi<\zeta^\alpha\le\pi,\quad
    \alpha=1,2,3
        \rbrace/\sim
\end{equation}
defines the 3-torus $\mathbb{F}^3$ as a compact 3-manifold with no boundary, the space formed by a solid rectangular box with each pair of opposite faces identified as the same surface (Fig. \ref{f:Ffz}, left). 
It may be thought of as gluing the opposite faces together\cite{thurston-1997} in the same way that rolling up a flexible rectangle and gluing its parallel edges constructs a 2-torus.
$\mathbb{F}^3$ is not bounded by the faces of the illustrated cube; the faces and all other points of the torus manifold are interior points.
Identifying each pair of opposite faces results in each set of four parallel edges being identified as the same edge, and all eight corners being identically one point, $(\pi,\pi,\pi)$.
The cube model's lines represent closed loops with zero curvature.
$\mathbb{F}^3$ paired with the Euclidean metric $g_{\alpha\beta}=\delta_{\alpha\beta}$ defines the metric space $(\mathbb{F}^3,\delta_{\alpha\beta})$, a \emph{flat} 3-torus, as the quotient space of $\mathbb{R}^3$ and the lattice $2\pi\mathbb{Z}^3$.
\label{s:Brho-too-small}
\label{s:nonsimply}
The shortest non-simply connected loops in $(\mathbb{F}^3,\delta_{\alpha\beta})$ are its coordinate lines.
There are three sets of coordinate lines in $(\mathbb{F}^3,\delta_{\alpha\beta})$: $(\zeta^1,c^2,c^3)$, $(c^1,\zeta^2,c^3)$, and $(c^1,c^2,\zeta^3)$ for constants $-\pi<c^i\le\pi$.
Each coordinate circle of flat 3-torus $(\mathbb{F}^3,\delta_{\alpha\beta})$ is both a closed loop and a straight line with no curvature. \label{s:entire-coord-circles}
Such a loop cannot be continuously shrunk to length less than $2\pi$ because in this metric space the loop has no disk interior in which to shrink.
\label{s:Brho}
Define a closed ball
$\mathbb{B}_p\subset(\mathbb{F}^3,\delta_{\alpha\beta})$ as the union of 
$\mathbb{S}_p$ and its interior.
Radius $p$ makes $\mathbb{B}_p$ too small to contain an entire coordinate circle. Thus, $\mathbb{B}_p$ can be continuously deformed to a point and is simply-connected.

\label{s:Sp}
Define a 2-sphere $\mathbb{S}_p\subset(\mathbb{F}^3,\delta_{\alpha\beta})$ centered at 
$(\pi,\pi,\pi)$ with radius $p<\pi$. See Fig. \ref{f:Ffz} (left). The inverted image of this sphere will serve later as the simply-connected boundary in Definition 1.
The 2-sphere $\mathbb{S}_p$ is a simply-connected sphere centered on the corner point. 
Fig. \ref{f:Ffz} (left) shows 
$\mathbb{S}_p$ in separate octants because in the flat 3-torus cube model no space exists outside the cube. 
Magenta labels ``S'' and ``N'' indicate simply- and nonsimply-connected volumes of space, discussed below.

\label{s:circleL}
The shortest straight line $L$ intersecting all of the coordinate lines in any one of the three sets of them in $(\mathbb{F}^3,\delta_{\alpha\beta})$ must have length $2\pi$ and must be any coordinate circle from one of the other 2 sets. Since $p<\pi$, the intersection $L\cap\mathbb{B}_p$ is too short to intersect all the coordinate circles in the full set. Coordinate circles in the full set which do not intersect $L$ exist wholly in the exterior of $\mathbb{B}_p$. Therefore, 
\begin{equation}\label{e:F3Bp}
    \mathbb{F}^3\setminus\mathbb{B}_p,
    \text{ the exterior of }\mathbb{S}_p
    \text{ is non-simply connected.}
\end{equation}

\label{s:FF3}
Equation (\ref{e:f}) is a higher-dimensional variation of Killing's flat torus mapping (\ref{e:killing}) without normalization. It converts from (triple) polar to Cartesian coordinates in $\mathbb{R}^6$, and for each set of $r^\alpha>0$ maps a unique flat 3-torus into 
$\mathbb{R}^6$.
Treating the $r^\alpha$ as coordinates with the $\zeta^\alpha$ we see that for any fixed $\sigma>0$ the mapping (\ref{e:f}) foliates $\mathbb{R}^6$ with flat 3-tori of all sizes and aspect ratios.
\small
\begin{equation}
  {\bf f}\left(\zeta^1,\zeta^2,\zeta^3,
   r^1,r^2,r^3\right)
  = 
  \left(
  \sigma+r^1\cos\zeta^1,
  r^1\sin\zeta^1,
  \sigma+r^2\cos\zeta^2,
  r^2\sin\zeta^2,
  \sigma+r^3\cos\zeta^3,
  r^3\sin\zeta^3 
  \right)
  \nonumber
\end{equation}
\normalsize
\small
\begin{equation}\label{e:f}
  r^\alpha\ge0,\quad
  \sigma>0,\quad
  -\pi<\zeta^\alpha\le\pi,\quad
  \alpha=1,2,3.
\end{equation}
\normalsize
We also formulated similar triple wormhole cases replacing $\sigma$ with three distinct translation distances in directions 1, 3, and 5, but omit them here for simplicity.

To isometrically embed the intrinsically flat unit-radius 3-torus 
$\mathbb{F}$ as a 3-manifold with extrinsic curvature in $\mathbb{R}^6$ we write
\begin{align}
    \label{e:fF3}
    {\bf f}\left(\mathbb{F}^3\right)
    &\coloneqq
    \left\{{\bf f}\left(\zeta^1,\zeta^2,\zeta^3,1,1,1\right):\left(\zeta^1,\zeta^2,\zeta^3\right)\in\mathbb{F}^3\right\},
\end{align}
Likewise we write 
${\bf f}\left(\mathbb{S}_p\right)$ and
${\bf f}\left(\mathbb{B}_p\right)$ to isometrically embed the 2-sphere and ball.

Solely to aid visualization, Fig. \ref{f:Ffz} (center) plots a surjection $\mathsf{P}{\bf f}\left(\mathbb{F}^3\right)$, (Fig. \ref{f:Ffz}, left) from $\mathbb{R}^6$ to $\mathbb{R}^3$, using
\begin{equation}\label{e:P}
    \mathsf{P}\coloneqq
    \begin{bmatrix}
      0 & 1 & 1 & 0 & 0 & 0 \\
      0 & 0 & 0 & 1 & 1 & 0 \\
      1 & 0 & 0 & 0 & 0 & 1
    \end{bmatrix}.
\end{equation}
Likewise, $\mathbb{B}_p$ and $\mathbb{S}_p$ are embedded as 
$\mathbf{f}\left(\mathbb{B}_p\right)$ and 
$\mathbf{f}\left(\mathbb{S}_p\right)$ respectively, (illustrated by the surface
$\mathsf{P}\mathbf{f}\left(\mathbb{S}_p\right)$ in Fig. \ref{f:Ffz}, center). 
Each of these three images is the next-higher dimensional analogue of a respective part of the 2-torus image in Fig. \ref{f:inv2Fboundary}. 
The surjected embedded sphere $\mathsf{P}\mathbf{f}(\mathbb{S}_p)$ viewed in this illustration is no longer spherically symmetric for reason analogous to that of a circle, which being drawn on a rectangular sheet of paper then rolled into a cylinder, no longer has circular symmetry in $\mathbb{R}^3$. 
The surjection $\mathsf{P}$ necessarily distorts some shapes and is for illustration only, playing no role in the proof.
The exterior of $\mathbf{f}\left(\mathbb{S}_p\right)$ embedded in $\mathbb{R}^6$ inherits non-simply connectedness from $\mathbb{F}^3\setminus\mathbb{B}_p$.
This is seen in Fig. \ref{f:Ffz} (center) where the blue coordinate lines intersect the embedded sphere while the orange anchor circles and gray extreme circles are exterior to $\mathbf{f}(\mathbb{S}_p)$.

\label{s:S5}
We invert (\ref{e:f}) in $\mathbb{R}^6$ through a 5-sphere centered at the origin
with radius
\begin{equation}\label{e:q}
    q\coloneqq
    \left|{\bf f}(0,0,0,\sigma,\sigma,\sigma)
    -{\bf f}(\pi,\pi,\pi,\sigma,\sigma,\sigma)\right|
    =\sigma\sqrt{12}.
\end{equation}
This inversion's conformal factor is
\small
\begin{equation}\label{e:lambda-z}
    \lambda\coloneqq
    \dfrac{q^2}
    {\left|{\bf f}\right|^2}
    =
    \frac{12}
    {\sum_{\alpha=1}^3
    \big(1+2(r^\alpha/\sigma)\cos\zeta^\alpha+(r^\alpha/\sigma)^2\big)}.
\end{equation}
\normalsize
Note the three symmetries of conformal factor $\lambda$:
\begin{equation}\label{e:lambda-sym}
    \lambda|_{\zeta^\beta=\phi} =
    \lambda|_{\zeta^\beta=-\phi}, \quad
    \text{ for each } \beta=\{1,2,3\}, 
    \text{ and }-\pi<\phi\le\pi.
\end{equation}

The scale factor $q$ is arranged so that the conformal factor is 1 at ${\bf \zeta}=(0,0,0)$ when all $r^\alpha=\sigma$. 
Mapping ${\bf z}$ below, like mapping ${\bf f}$ (\ref{e:f}), is a foliation covering $\mathbb{R}^6$ with inverted flat 3-tori because spherical inversion is homeomorphic.
\begin{align}
  \label{e:z}
  {\bf z}\left(\zeta^1,\zeta^2,\zeta^3,
   r^1,r^2,r^3\right)
  \coloneqq \lambda{\bf f}(\zeta^1,\zeta^2,\zeta^3,r^1,r^2,r^3)
\end{align}
with diagonal spatial metric, 
\begin{equation}\label{e:z3metric}
    ds^2=
    \left(
    \frac{12}
    {\sum_{\alpha=1}^3
    \big(1+2(r^\alpha/\sigma)\cos\zeta^\alpha+(r^\alpha/\sigma)^2\big)}
    \right)
    \sum_{\alpha=1}^3
    \left(r^\alpha d\zeta^\alpha\right)^2,\quad
    r^\alpha>0.
\end{equation}

A mapping with greater symmetry is constructed by noticing that the denominator of (\ref{e:lambda-z}) vanishes only at one point, where all 
$\zeta^\alpha=\pi$, and only for the case where all 
$r^\alpha=\sigma$. Therefore, the only unbounded case of inverted 3-torus in (\ref{e:z}) is
\begin{align}
  \label{e:w}
  {\bf w}\left(\zeta^1,\zeta^2,\zeta^3\right)
  \coloneqq
  {\bf z}\left(\zeta^1,\zeta^2,\zeta^3,\sigma,\sigma,\sigma\right)
  =
  \frac{6{\bf f}\left(\zeta^1,\zeta^2,\zeta^3,\sigma,\sigma,\sigma\right)}
  {3+\cos\zeta^1+\cos\zeta^2+\cos\zeta^3}
  ,\quad\mathbf{\zeta}\ne(\pi,\pi,\pi),
\end{align}
with isothermal spatial metric, 
\begin{equation}\label{e:w3metric}
  ds^2=
  \left(\dfrac{6\sigma}{3+\cos\zeta^1+\cos\zeta^2+\cos\zeta^3}\right)^2
  \left(
  (d\zeta^1)^2
  +(d\zeta^2)^2
  +(d\zeta^3)^2\right),
\end{equation}
and conformal factor (\ref{e:lambda-z}) simplifies to
\begin{equation}\label{e:lambda-w}
  \lambda
  =
  \dfrac{6}{3+\cos\zeta^1+\cos\zeta^2+\cos\zeta^3}.
\end{equation}

\label{s:proof1conclusion}
Here we show that (\ref{e:w}) and (\ref{e:w3metric}) describe the spatial submanifold of a spacetime triple wormhole.
Mappings (\ref{e:z}) and (\ref{e:w}) preserve the triply-connected 3-torus spatial topology because each step of the transformation from $\mathbb{F}^3$ (\ref{e:f3}) to ${\bf w}$ (\ref{e:w}) is homeomorphic.
The boundary sphere's inverted image ${\bf w}(\mathbb{S}_p)$ inherits simply connectedness from $\mathbb{S}_p$. The interior of ${\bf w}(\mathbb{S}_p)$ in Fig. \ref{f:Ffz} (right) is the inverted image of the exterior of $\mathbb{B}_p$, so it inherits non-simply connectedness from $\mathbb{F}^3\setminus\mathbb{B}_p$ (\ref{e:F3Bp}).
The exterior of  ${\bf w}(\mathbb{B}_p)$ inherits simply connectedness from $\mathbb{B}_p$. Therefore, the non-simply connected interior volume of
${\bf w}(\mathbb{S}_p)$ taken as the $\Sigma$ in Definition 1 bounded by the simply-connected boundary $\partial\Sigma={\bf w}(\mathbb{S}_p)$ satisfies the spatial properties of Definition 1. 

We append a time coordinate $\zeta^0=t$ and satisfy the time property of Definition 1 by placing the spatial manifold (\ref{e:w3metric}) in a synchronous reference frame, as Misner\cite{misner-1960} did in 1960 with the spatial 3-manifold, $S^2\times S^1$ which can be identified as a Dupin hypercyclide\cite{pinkall-1985B}.
\begin{equation}\label{e:g0mu}
   g_{\mu0}=-\delta_{\mu0}.
\end{equation}
Condition (\ref{e:g0mu}) defines a ``synchronous frame'' described in more detail by
\citet[pp. 290-295]{landau-lifshitz-1971}.
From this with (\ref{e:w3metric}) we obtain the triple wormhole spacetime metric.
\begin{equation}\label{e:w4metric}
  ds^2
  =-dt^2+
  \left(\dfrac{6\sigma}{3+\cos\zeta^1+\cos\zeta^2+\cos\zeta^3}\right)^2
  \left(
  (d\zeta^1)^2
  +(d\zeta^2)^2
  +(d\zeta^3)^2\right).
\end{equation}
$\square$

\subsection{\textit{n}-neck wormholes in \textit{n}+1-spacetimes}
\label{s:n-neck}
For $n>3$, $n$-neck spacetime wormholes can be geometrically constructed for hypothetical $n$+1-spacetimes embedded in $\mathbb{R}^{2n+1}$ in like manner as above by directly expanding the number of dimensions in Eqs.\ (\ref{e:f3})-(\ref{e:w3metric}), \textit{e.g.,} an $n$-neck wormhole analogous to (\ref{e:w}) in $n$ spatial dimensions is described by the spatially isothermal
\begin{equation}\label{e:metric-n}
    ds^2=-dt^2+
    \frac{4n^2\sigma^2}
    {\left(n+\cos\zeta^1+\cos\zeta^2
    +\cdots+
    \cos\zeta^n\right)^2}
    \left((d\zeta^1)^2+(d\zeta^2)^2
    +\cdots+
    (d\zeta^n)^2\right).
\end{equation}
This can be found by following the same construction as above, replacing 3 by $n$,
$\mathbb{R}^6$ by $\mathbb{R}^{2n}$, 
2-sphere by $(n-1)$-sphere, 
5-sphere by $(2n-1)$-sphere, 
expanding sequences and vectors from 3 or 6 terms to $n$ or $2n$ terms, 
and so on.

\subsection{Diagonal stress-energy tensor}
\label{s:proof-diagonal}
Next we show that the triple wormhole $3+1$-spacetime diagonal coordinate metric
\begin{equation}\label{e:wmetric0123}
    ds^2= -dt^2
    +\lambda^2
    \sum_{\alpha=1}^3
    \left(r^\alpha d\zeta^\alpha\right)^2,\quad
    r^\alpha>0,
\end{equation}
formed by placing spatial manifold (\ref{e:z}) in a synchronous frame (\ref{e:g0mu}) generates a diagonal stress-energy tensor. 

\label{s:fhexads}
Extending the tetrad formalism to $\mathbb{R}^{6+1}$,
in place of the three spatial local basis vectors which Cartan\cite{cartan-2001} called \emph{orthogonal trihedra}, we assign a pre-inversion \emph{hexad} (\ref{e:APXfhexad}) or set of six simple and sparse spatial orthonormal local basis vectors
$\boldsymbol{\alpha}_m=\boldsymbol{\alpha}_m(\zeta^1,\zeta^2,\zeta^3,r^1,r^2,r^3)$ 
at each point 
${\bf f}={\bf f}(\zeta^1,\zeta^2,\zeta^3,r^1,r^2,r^3)\in\mathbb{R}^6$.
We assign a post-inversion \emph{hexad} of orthonormal local basis vectors 
 $\boldsymbol{\beta}_m=\boldsymbol{\beta}_m(\zeta^1,\zeta^2,\zeta^3,r^1,r^2,r^3)$ 
at each point 
${\bf z}={\bf z}(\zeta^1,\zeta^2,\zeta^3,r^1,r^2,r^3)\in\mathbb{R}^6$,
such that each $\zeta^i$ coordinate circle through ${\bf z}$ has one \emph{tangent} hexad spatial vector $\boldsymbol{\beta}_a$ and one \emph{normal} hexad spatial vector $\boldsymbol{\beta}_z$, 
normal and tangent respectively to the inverted embedded flat 3-torus having fixed $r^i$ in (\ref{e:z}). Each $\boldsymbol{\beta}_m$ is equivalent to the respective $\boldsymbol{\alpha}_m$ spherically inverted through the same $\mathbb{S}_q^5$ and then re-normalized to unit length, using a Householder reflection (\ref{e:APXH}). Thus, each hexad $\boldsymbol{\beta}_m$ has the same normal or tangent relationships with manifold ${\bf z}$ as the corresponding $\boldsymbol{\alpha}_m$ has with ${\bf f}$.
Hexads $\boldsymbol{\alpha}_m$ and $\boldsymbol{\beta}_n$ are normalizations of
$\partial{\bf f}/\partial{x^m}$ and
$\partial{\bf z}/\partial{x^n}$.
The resulting specific hexad values $\boldsymbol{\beta}_m$ on ${\bf z}$ are at (\ref{e:APXzhexad2}).
To form a heptad $\boldsymbol{\gamma}_m$ of seven orthonormal local basis vectors for $\mathbb{R}^{6+1}$ with spatial submanifold mapped by (\ref{e:z}), we prepend a zero-indexed component equal to zero for each of the six $\boldsymbol{\beta}_m$ for the time coordinate $x^0$ and include the unit-length timelike tangent vector 
$\boldsymbol{\gamma}_0=(1,0,0,0,0,0,0)$,
orthogonal to each of the other six spacelike heptad vectors.

Coordinate frame vector indices in this paper are identified by Greek letters, with early letters $\alpha$, $\beta$ reserved only for nonzero spatial directions $1, 2$ or $3$. Heptad frame indices are Latin letters identifying which heptad vector (not which of its components).
Early letters $a,b,...$ can be $0,1,2,$ or $3$ used for the tangent heptad vectors; $i$ and $j$ are reserved for spatial directions $1,2,$ or $3$ only;
later letters $...\;,y,z$ are $4,5,$ or $6$ used for the normal heptad vectors; 
middle letters $k,l,m,n,...$ can represent any heptad vector.
The four $\boldsymbol{\gamma}_a$ span
the \textit{tangent subspace} of the local heptad frame. The three $\boldsymbol{\gamma}_z$ span the heptad's \textit{normal subspace}. 

\label{s:zhexads}
To calculate heptad \emph{tensors} we choose orthonormal heptad vectors $\boldsymbol{\gamma}_m$ that produce a Minkowski \emph{heptad metric},
\begin{equation}\label{e:heptadmetric}
    \gamma_{mn}=\eta_{mn}.
\end{equation}
More hexad/heptad details are in Appendix \ref{s:APXTgeom}.

In general, the siebenbein 
${e^m}_{\mu}$ 
defined by 
$\mathbf{e}_\mu={e^m}_{\mu}\boldsymbol{\gamma}_m$ 
\citet[Eq.\ (11.4)]{hamilton-grbook} 
transform from the heptad frame to the coordinate frame and relate the coordinate and heptad metrics by 
\citet[Eq.\ (11.8)]{hamilton-grbook}, 
\begin{equation}\label{e:heptadmetric2}
    g_{\mu\nu}=\gamma_{mn}
    {e^m}_{\mu}{e^n}_{\nu}.\quad
\end{equation}
The set of siebenbein $\{({e^0}_\mu,\dots,{e^6}_\mu)\}_{\mu=0}^6$ is a locally Minkowskian basis as used in
\citet[Eqs.\ (3.2.14) and (12.5.1)]{weinberg-1972}.
Since our $\gamma_{mn}$ is Minkowski  and our $g_{\mu\nu}$ is diagonal, we have diagonal siebenbein with entries
\begin{equation}\label{e:siebenbein}
  \left({e^0}_{0},\cdots,{e^6}_{6}\right)
  =
  \left(1,\lambda r^1,\lambda r^2,\lambda r^3,
  \lambda,\lambda,\lambda
  \right).
\end{equation}
The inverse siebenbein ${e_n}^{\nu}$ are defined as the matrix inverse of the siebenbein 
(\citet[Eq.\ (11.5)]{hamilton-grbook}).
Along the heptad axes $\boldsymbol{\gamma}_n$, the diagonal ${e_n}^{\nu}$ simplify the directional derivatives 
(\citet[Eq.\ (11.30)]{hamilton-grbook}), 
\begin{equation}\label{e:dd}
  \partial_n
  ={e_n}^{\nu}
  \dfrac{\partial}{\partial x^\nu}\quad
  ,\quad x^\nu\in
  \left\lbrace t,\zeta^1,\zeta^2,\zeta^3, r^1,r^2,r^3\right\rbrace
\end{equation}
because for each $n$, only the $\nu=n$ case of
${e_n}^{\nu}$ is nonzero.
The \emph{inverse siebenbein derivative}, $d_{lmn}\coloneqq
\boldsymbol{\gamma}_{lk}{e_m}^{\mu}
{e_n}^{\nu}
\dfrac{\partial {e^k}_{\mu}}{\partial x^\nu}$,
not a tensor,
(\citet[Eq.\ (11.33)]{hamilton-grbook}) 
is simplified by (\ref{e:dd}) and the diagonal siebenbein (\ref{e:siebenbein}) to
\begin{equation}\label{e:dlmn}
    d_{lmn}
    =\boldsymbol{\gamma}_{ll}
    {e_m}^{\mu}
    \partial_n {e^l}_{\mu}\quad
    \text{(summation over }\mu\text{ only),}
\end{equation}
which must vanish when $l\ne m$ or any index is zero. See specific $d_{lmn}$ values at (\ref{e:APXdlmn1}).

\label{s:hexadconnection}
We denote the heptad-frame connection by $\Gamma_{lmn}$ (not a tensor, and with Latin indices not a Christoffel symbol). In general, the $\Gamma_{lmn}$ are antisymmetric in their first two indices by their definition structure,
(\citet[Eq.\ (11.54)]{hamilton-grbook})
\begin{equation}\label{e:heptadconnection}
    \Gamma_{lmn}:=-\Gamma_{mln}:=\frac{1}{2}
    \left(d_{lnm}-d_{lmn}+d_{mln}-d_{mnl}+d_{nlm}-d_{nml}\right)
\end{equation} 
The sparsity of our $d_{lmn}$ (\ref{e:dlmn}) simplifies (\ref{e:heptadconnection}) to
\begin{equation}\label{e:heptadconnection2}
    \Gamma_{lml}=-\Gamma_{mll}=d_{llm},\quad 
    l\ne m,l\ne0,m\ne0,\quad
    \text{ all other }\;\Gamma_{lmn}=0.
\end{equation} 
\label{s:Kazm}
\indent
The extrinsic curvature $K_{azn}$ in general is defined 
(\citet[Eqs.\ (17.191)-(17.195)]{hamilton-grbook})
to be the heptad-frame connections with first two indices from opposite subspaces ($a\in\{0,1,2,3\}, z\in\{4,5,6\}$), otherwise zero. Its complement is the restricted heptad-frame connection $\hat{\Gamma}_{lmn}$,
\begin{equation}\label{e:Kazn}
  K_{azn}\coloneqq-K_{zan}\coloneqq\Gamma_{azn};\quad
  K_{abn}\coloneqq0;\quad K_{yzn}\coloneqq0;\quad
  \hat{\Gamma}_{lmn}\coloneqq\Gamma_{lmn}-K_{lmn}.
\end{equation}
In our case (\ref{e:heptadconnection2}) simplifies the extrinsic curvature to
\begin{equation}\label{e:Kizi}
    K_{izi}=-K_{zii} = d_{iiz},\quad
    K_{ziz}=-K_{izz} = d_{zzi},\quad
    \text{ all other }\;K_{lmn}=0.
\end{equation}
If no two indices are equal, then $K_{lmn}=0$. 
Thus, we call $K_{lmn}$ diagonal.
If any index is zero, then $K_{lmn}=0$. Combining this fact and the Minkowski heptad metric (\ref{e:heptadmetric}) we obtain 
${K^z}_{cb}=K_{zcb}$ used in Gauss' equation (\ref{e:gauss}) below.
From (\ref{e:APXdlmn1}) we obtain
\begin{equation}\label{e:Kz}
    K_z={K_{za}}^a
    =-{d^a}_{az}
    =\dfrac{\cos\zeta^i}
    {2\sigma}
    +\dfrac{r^i}
    {2\sigma^2}
    -\dfrac{1}{\lambda r^i}, 
    \text{ where }i=z-3,
    \text{ summation over }\textit{a}
    \text{ only}.
\end{equation}

We are interested in the restricted Riemann tensor
$\hat{R}_{abcd}$,
defined by (\ref{e:APXhatRabcd}), which describes the intrinsic curvature of the 3+1-spacetime containing the triple wormhole, presently embedded for mathematical convenience in
$\mathbb{R}^{6+1}$ spacetime.
Let $R_{klmn}$ denote the heptad Riemann tensor (\ref{e:APXRklmn}) of the parent manifold which is
$\mathbb{R}^{6+1}$ spacetime. The relation between them is Gauss' equation
(\citet[Eq.\ (17.197a)]{hamilton-grbook})
decomposing $R_{abcd}$ into intrinsic and extrinsic curvatures.
\begin{equation}\label{e:gauss}
    R_{abcd} = 
    \hat{R}_{abcd}
    + {K^z}_{cb} K_{zda} - {K^z}_{ca} K_{zdb}.
\end{equation}
Since the spherical inversion is only a coordinate change of flat $\mathbb{R}^6$, the left-hand side heptad Riemann tensor $R_{abcd}=0$,
implying that the embedded manifold's intrinsic curvature is
\begin{equation}\label{e:hatRabcd}
    \hat{R}_{abcd} 
    = -{K^z}_{cb} K_{zda} 
    + {K^z}_{ca} K_{zdb}.
\end{equation}
Since by definition $z\ne b$ and $z\ne c$, we know from (\ref{e:Kizi}) that our ${K^z}_{cb}$ must vanish in (\ref{e:hatRabcd}) when $b\ne c$.
The sparsity and symmetry of $K_{lmn}$ (\ref{e:Kizi}) simplifies (\ref{e:hatRabcd}) to
\begin{align}
    \nonumber
    \hat{R}_{ijij} 
    =&
    -\hat{R}_{ijji} = {K^z}_{ii} K_{zjj}
    =
    -\dfrac{\frac{1}{r^i}\cos\zeta^i+\frac{1}{r^j}\cos\zeta^j}{6\sigma\lambda}
    -\dfrac{\sin^2\zeta^1+\sin^2\zeta^2+\sin^2\zeta^3}{36\sigma^2},\\
    &
    \label{e:hatRabcd1}
    \text{ where }i\ne j;
    \text{ all other } \hat{R}_{abcd}=0.
\end{align}
The structure of Eq. (\ref{e:hatRabcd1}) and the symmetries (\ref{e:lambda-sym}) of the conformal factor impart three symmetries also to the intrinsic curvature:
\begin{equation}\label{e:hatRijij-sym}
    \hat{R}_{ijij}|_{\zeta^k=\phi} =
    \hat{R}_{ijij}|_{\zeta^k=-\phi}, \quad
    \text{ for each } k=\{1,2,3\}, 
    \text{ and }-\pi<\phi\le\pi.
\end{equation}

Contracting the general case (\ref{e:hatRabcd}) and using the antisymmetry ${{K_z}^b}_a=-{K^b}_{za}$ gives us the embedded manifold's Ricci tensor,
\begin{equation}\label{e:hatRac}
    \hat{R}_{ac} = 
    \hat{R}_{abc}^{\;\;\;\;\;\;b} =
    {K^z}_{cb} {K^b}_{za} + {K^z}_{ca} K_z,
\end{equation}
which in our case by (\ref{e:Kizi}) vanishes when $a\ne c$ and also when either is zero. Thus, our Ricci tensor is diagonal. 
Substituting (\ref{e:hatRabcd1}) into (\ref{e:hatRac}) gives us
\begin{align}\label{e:hatRii}
    \hat{R}_{00}
    =0;\;
    \hat{R}_{ii} 
    = 
    \hat{R}^{\;\;j}_{\,i\;ij} =
    {K^z}_{ii} {K^i}_{zi}
    +{K^z}_{ii} K_z,\;
    \text{no summation over $i$.}
\end{align}
Equations (\ref{e:Kz}) and (\ref{e:hatRabcd1}) simplify the restricted Ricci tensor to 
\begin{align}\label{e:hatRii2}
    \hat{R}_{00}
    =0;\;
    \hat{R}_{ii} 
    =
    -\dfrac{\frac{1}{r^1}\cos\zeta^1+\frac{1}{r^2}\cos\zeta^2+\frac{1}{r^3}\cos\zeta^3+\frac{1}{r^i}\cos\zeta^i}{6\sigma\lambda}
    -\dfrac{\sin^2\zeta^1+\sin^2\zeta^2+\sin^2\zeta^3}{18\sigma^2}.
\end{align}

In summary, since the extrinsic curvature tensor (\ref{e:Kazn}) is diagonal, the restricted Ricci heptad tensor (\ref{e:hatRac}) is diagonal. This and the diagonal heptad metric (\ref{e:heptadmetric}) also make the Einstein tensor diagonal. The stress-energy tensor is diagonal because it equals the Einstein tensor times a constant.
$\square$
%
%
%
%

\subsection{Asymptotic flatness}\label{s:flatness}
We see that the inverted flat 3-torus spacetime metric (\ref{e:w4metric}) is asymptotically flat because the limit of the heptad frame restricted curvature tensor vanishes in the far field:
\begin{align}
    \lim_{\zeta\to(\pi,\pi,\pi)}
    \hat{R}_{abab}(\boldsymbol{\zeta}) &
    =0.
\end{align}
More specifically, from the series expansion in $h$, for small radius-$h$ sphere around the inversion point, \textit{i.e.} 
$\boldsymbol{\zeta}=(\pi+h\cos{\theta}\sin{\phi},
\pi+h\sin{\theta}\sin{\phi},
\pi+h\cos{\phi})$
we obtain
\footnotesize
\begin{align}
    \nonumber
    \hat{R}_{1212}= -\hat{R}_{1221}
    &
    =\dfrac{\cos^4{\phi}
    -\cos^2{\phi}\sin^2{\phi}
    -2\cos^2{\theta}\sin^2{\theta}\sin^4{\phi}}
    {144\sigma^2}h^4
    +O[h]^6,
    \\
    \nonumber
    \hat{R}_{1313}= -\hat{R}_{1331}
    &
    =
    -\dfrac{
    (\cos^2{\phi}-\sin^2{\theta}\sin^2{\phi})
    \sin^2{\theta}
    +(2\cos^2{\phi}
    +\sin^2{\theta}\sin^2{\phi})
    \cos^2{\theta}}
    {144\sigma^2}h^4\sin^2{\phi}
    +O[h]^6,
    \\
    \nonumber
    \hat{R}_{2323}= -\hat{R}_{2332}
    &
    =-\dfrac{
    (2\cos^2{\phi}+\cos^2{\theta}\sin^2{\phi})
    \sin^2{\theta}
    +(\cos^2{\phi}-\cos^2{\theta}\sin^2{\phi})
    \cos^2{\theta}
    }
    {144\sigma^2}h^4\sin^2{\phi}
    +O[h]^6,
    \\
    \label{e:hatR1212}
    & \quad\text{ with all other } \hat{R}_{abcd}=0,
\end{align}
\normalsize
where Eqs.\ (\ref{e:w4metric}) and (\ref{e:APXChristoffel}) imply coordinate and proper times $t$ and $\tau$ are equal up to a shift and scale.

Note that by Eq.\ (\ref{e:w}), the distance from the manifold center is $O(h^{-1})$; and by \citet[Eq.\ 31.4b]{mtw}, the Riemann curvature of a Schwarzschild mass $M$ in our coordinates would be $O(Mh^3)$. Since the triple-wormhole curvature is zero to order $h^3$, in this sense its total mass  is zero in the synchronous frame. 

\subsection{Geodesics}
\label{s:geodesics}
\begin{figure}
   \begin{center}
      \includegraphics[width=1.0\linewidth]{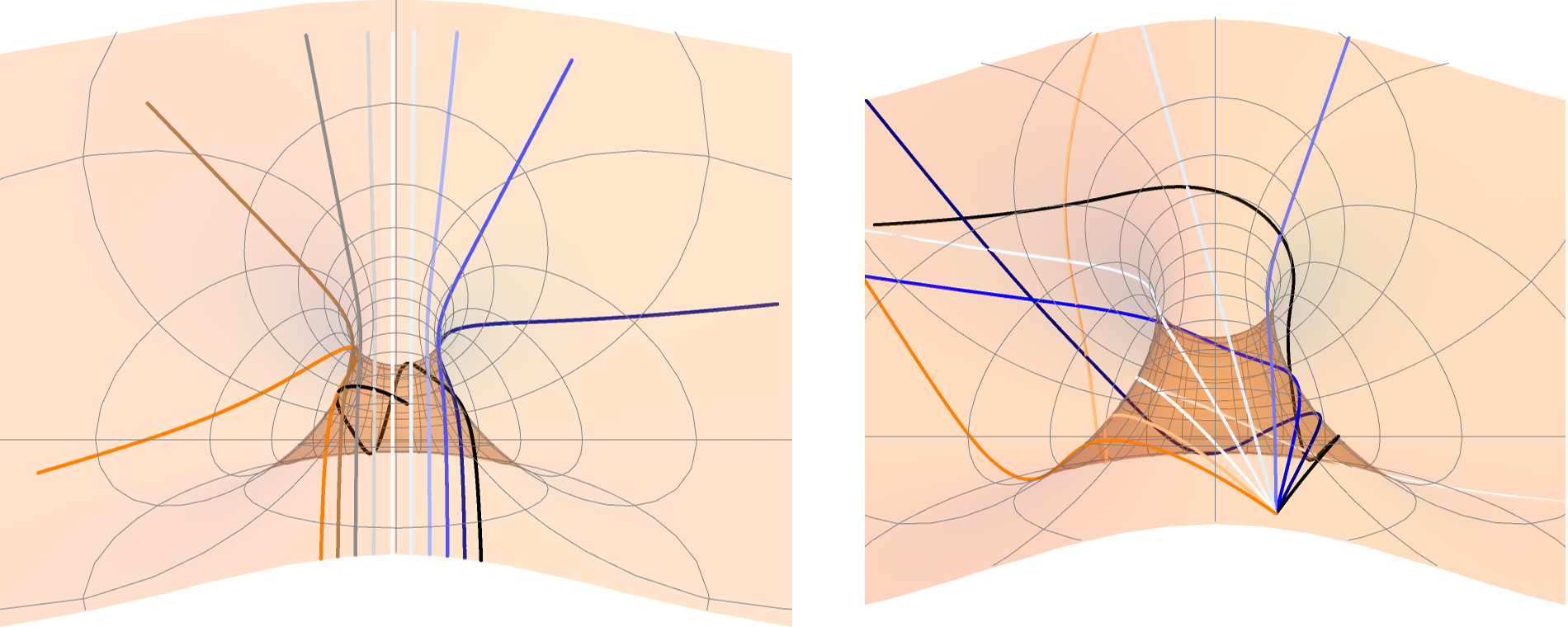}
   \end{center}
   \caption{\footnotesize
   Two sets of geodesics near and through one neck of the triple wormhole. Each surface is a PDC representing a spherically inverted face of the the flat 3-torus box model (Fig. \ref{f:Ffz}, left), e.g., ${\bf w}(\zeta^1,\zeta^2,\pi)$. (See Appendix \ref{s:APXPDC}.
   }
   \label{f:geodesics}
\end{figure}
In the coordinate frame, the equation of motion (\citet[Eq.\ 3.2.3]{weinberg-1972}) is
\begin{equation}
  \label{e:eom}
    \dfrac{d^2\zeta^\lambda}{dt^2}
    =-\Gamma^\lambda_{\mu\nu}
    \dfrac{d\zeta^\mu}{dt}
    \dfrac{d\zeta^\nu}{dt}.
\end{equation}
In this section we calculate geodesics on the spacetime triple wormhole with metric (\ref{e:w4metric})
for which the affine connection (\ref{e:APXChristoffel}) coefficients simplify to
\begin{align}
    \nonumber
    \Gamma^\alpha_{\alpha\alpha}=
    \Gamma^\beta_{\alpha\beta}=
    \Gamma^\beta_{\beta\alpha}=
    -\Gamma^\alpha_{\beta\beta}=
    \dfrac{\sin\zeta^\alpha}{3+\cos\zeta^1+\cos\zeta^2+\cos\zeta^3},\quad
    \alpha\ne\beta,\quad
    \alpha,\beta\in\{1,2,3\},
    \\
    \label{e:affine}
    \text{and all other }
    \Gamma^\lambda_{\mu\nu}=0,\quad
    \lambda,\mu,\nu\in\{0,1,2,3\}.
\end{align}

The time lines for fixed coordinate points in this spacetime are geodesics. 
Since the affine connection vanishes for any zero index,
objects at rest relative to the spatial coordinates have zero geometric acceleration. Landau and Lifshitz\cite{landau-lifshitz-1971} describe this counterintuitive result as a direct, inevitable consequence of spacetime metric condition (\ref{e:g0mu}).

In general, trajectories following the spatial coordinate lines are not geodesics, but notable exceptions include the central paths through the three necks:  ${\bf w}(\zeta^1,\pi,\pi)$, 
${\bf w}(\pi,\zeta^2,\pi),$ and 
${\bf w}(\pi,\pi,\zeta^3)$, illustrated by the blue lines in Fig. \ref{f:Ffz}. 
This mapping (\ref{e:w}) inherited its spatial metric from $\mathbb{R}^6$, so we know the three central neck paths are geodesics because their embedded images are straight lines in 
$\mathbb{R}^6$, \textit{e.g.,} 
\begin{equation}\label{e:neck1}
    {\bf w}(\zeta^1,\pi,\pi)=6\sigma\left(
    1,\tan\frac{\zeta^1}{2},0,0,0,0\right).
\end{equation}
Given a 4-velocity ${\bf u}=\left(u^0,u^1,0,0\right)$ along this line (\ref{e:neck1}) in the restricted $3+1$-spacetime, the coordinate acceleration calculated by (\ref{e:eom}) simplifies to
\begin{equation}\label{e:accel-neck1}
    \dfrac{d^2\boldsymbol{\zeta}}{dt^2}
    =
    -\left(0,
    \left(u^1\right)^2
    \tan\dfrac{\zeta^1}{2},
    0,0\right),
\end{equation}
At an arbitrary starting point ${\bf w}(\zeta^1(0),\pi,\pi)$  on line ${\bf w}(\zeta^1,\pi,\pi)$, given an initial colinear velocity, the equation of motion (\ref{e:eom}) with up-index
$\lambda=1$ indicates nonzero coordinate acceleration only in the directions $\pm\zeta^1$, confirming that the
${\bf w}(\zeta^1(0),\pi,\pi) $coordinate line is a geodesic.
On this geodesic, nonzero coordinate acceleration is always toward the neck midpoint
${\bf w}(0,\pi,\pi)$.
To solve the equation of motion on this line, we note that
Eq. (\ref{e:affine}) with $\zeta^2=\zeta^3=\pi$ implies $\Gamma^0_{\alpha\beta}=\Gamma^2_{\alpha\beta}=\Gamma^3_{\alpha\beta}=0$. Therefore Eq. (\ref{e:eom}) becomes 
\begin{align}
\frac{d^2\zeta^1}{dt^2}&=-\left(\frac{d\zeta^1}{dt}\right)^2\tan\frac{\zeta^1}2\\
\Rightarrow\quad\zeta^1(t)&=2\arctan\frac{\dot\zeta^1(0)t+\sin\zeta^1(0)}{1+\cos\zeta^1(0)}.
\end{align}

Similarly, geodesics lie on the coordinate circles
${\bf w}(\zeta^1,0,0)$, 
${\bf w}(0,\zeta^2,0),$ and 
${\bf w}(0,0,\zeta^3)$, illustrated by the gray circles in Fig. \ref{f:Ffz}, \emph{e.g.}, for the same initial 4-velocity ${\bf u}$ on circle ${\bf w}(\zeta^1,0,0)$
the equation of motion (\ref{e:eom}) indicates nonzero coordinate acceleration only in the directions $\pm\zeta^1$, confirming that the
${\bf w}(\zeta^1,0,0)$ coordinate circle is a geodesic.
For geodesics on circle ${\bf w}(\zeta^1,0,0)$, nonzero coordinate acceleration is always toward the point
${\bf w}(\pi,0,0)$,
and test particles at the points
${\bf w}(\pi,0,0)$ and ${\bf w}(0,0,0)$
have zero acceleration whether in motion or not.

Spanning the narrowest section of each of the three wormhole necks are two of the six equal minimum-length coordinate circles
, \textit{e.g.,} spanning neck 1 are coordinate circles 
${\bf w}\left(0,\zeta^2,\pi\right)$ and 
${\bf w}\left(0,\pi,\zeta^3\right)$. These and the corresponding pairs of circles for necks 2 and 3 are the other geodesic coordinate line exceptions because, given a velocity colinear with a vector tangent to any of these circles, equation of motion (\ref{e:eom}) produces nonzero acceleration only along the same tangent.

Next, consider the coordinate surface 
${\bf w}(\zeta^1,\zeta^2,\pi)$ corresponding in the pre-inversion model to one face of the cube in Fig. \ref{f:Ffz} (left). At any point on this surface, tangent velocity vectors produce $\dfrac{d^2\zeta^3}{dt^2}=0$ in equation of motion (\ref{e:eom}). Thus geodesic paths on this surface never leave the surface. Similarly for surfaces 
${\bf w}(\zeta^1,\pi,\zeta^3)$, 
${\bf w}(\pi,\zeta^2,\zeta^3)$, 
${\bf w}(\zeta^1,\zeta^2,0)$, 
${\bf w}(\zeta^1,0,\zeta^3)$, and 
${\bf w}(0,\zeta^2,\zeta^3)$.
A rigid rotation (\ref{e:APXrot}) in $\mathbb{R}^6$ shows that the surface ${\bf w}(\zeta^1,\zeta^2,\pi)$ is a PDC and can be isometrically embedded in $\mathbb{R}^3$. Accordingly, Fig. \ref{f:geodesics} shows a selection of geodesic paths on this section of the triple wormhole. 

Point ${\bf w}(0,\pi,\pi)$ is the midpoint of the central path through neck 1. We showed that the coordinate lines 
${\bf w}(0,\zeta^2,\pi)$ and 
${\bf w}(0,\pi,\zeta^3)$ are geodesic circles, and that every stationary coordinate point is on a timelike geodesic. Therefore, if the neck was large enough and if the stipulated negative energy density within was harmless and transparent, then a person at the midpoint of neck 1, looking in the direction along either of these two coordinate lines orthogonal to the neck direction would see the image of the back of their own head at some distance, courtesy of the torus topology.

\subsection{Satisfying the geometric definition of spacetime wormhole}
\label{s:geometric-def}
The visual evidence for the two-dimensional $\zeta^\alpha =\pi$ sections (each identical to Fig. \ref{f:dupin}) of mapping (\ref{e:w}) embedded in Euclidean space suggest that the $2$-surface
${\bf w}(0,\zeta^2,\zeta^3)$ is a minimal-area surface as required by geometric wormhole throat definition in section III of Hochberg-Visser\cite{hochberg-visser-1997}. On this surface, point 
${\bf w}(0,\pi,\pi)$ is the midpoint of the straight-line central geodesic path
${\bf w}(\zeta^1,\pi,\pi)$ through neck 1.

Wormhole necks have no walls. For each constant
$-\pi<C<\pi$, the two-dimensional coordinate surface
${\bf w}(C,\zeta^2,\zeta^3)$ orthogonal to the neck geodesic above is closed and has no boundary.
Each such 2-surface has finite area because the conformal factor
$\lambda$ (\ref{e:lambda-w}) is finite at every point on it.
Each such 2-surface is an elliptical Dupin cyclide, diffeomorphic to a 2-torus, because it is the spherically inverted internal orthogonal slice
${\bf f}(C,\zeta^2,\zeta^3)$ through the embedded original flat 3-torus illustrated by the cube in Fig. (\ref{f:Ffz}). The surface formed by $C=\pi$
(a square face of the cube in Fig. \ref{f:Ffz})
when spherically inverted is the unbounded parabolic Dupin cyclide of Fig. \ref{f:dupin}.

By \emph{throat} we mean the minimal-area constant-time 2-surface orthogonal to the wormhole neck geodesic as in Hochberg and Visser\cite{hochberg-visser-1997} section III. We locate the triple wormhole's three $2$-surface wormhole throats and prove that they satisfy Hochberg-Visser's 
wormhole throat definition and their strong flare-out condition.
Using the 
isothermal $\boldsymbol{\zeta}$-coordinates of mapping (\ref{e:w}) instead of their stricter requirement of Gaussian normal coordinates we now offer a simpler minimal-area proof. 

First, we examine the $2$-surface orthogonal section
${\bf w}(0,\zeta^2,\zeta^3)$ associated with the geodesic
${\bf w}(\zeta^1,\pi,\pi).$
Here we will show that the two-surface
${\bf w}(0,\zeta^2,\zeta^3)$ at the midpoint of neck $1$ is a locally minimal-area surface orthogonal to the neck geodesic
${\bf w}(\zeta^1,\pi,\pi)$.
The symmetry (\ref{e:hatRijij-sym}) gives us symmetric intrinsic curvatures on opposite sides of the $2$-surface
${\bf w}(0,\zeta^2,\zeta^3)$,
\begin{equation}\label{e:throat-sym}
    \hat{R}_{ijij}(\delta,\zeta^2,\zeta^3) =
    \hat{R}_{ijij}(-\delta,\zeta^2,\zeta^3).
\end{equation}
From Eq. (\ref{e:lambda-w}) we obtain the symmetric conformal factor inequality
\begin{equation}\label{e:throat=lambda}
    \lambda(\delta,\zeta^2,\zeta^3) =
    \lambda(-\delta,\zeta^2,\zeta^3)>
    \lambda(0,\zeta^2,\zeta^3).
\end{equation}
Since in mapping ${\bf w}$ (\ref{e:w}) the $\boldsymbol{\zeta}$-coordinates are orthogonal everywhere, each tangent vector in the 
$\pm\zeta^1$ direction at any point on
${\bf w}(0,\zeta^2,\zeta^3)$ is orthogonal to this $2$-surface.
Metric (\ref{e:z3metric}), inherited from Euclidean
$\mathbb{R}^6$, gives us, for any choice of four small
$\delta_\alpha\ne0$, an inequality between the areas of two differential quadrilaterals on and near the $2$-surface
${\bf w}(0,\zeta^2,\zeta^3)$:
\begin{align}
    \nonumber
    \delta A 
    = 
    \left|{\bf w}(\delta_1,\zeta^2+d\zeta^2,\zeta^3)
    -{\bf w}(\delta_2,\zeta^2,\zeta^3)\right|
    \cdot
    \left|{\bf w}(\delta_3,\zeta^2,\zeta^3+d\zeta^3)
    -{\bf w}(\delta_4,\zeta^2,\zeta^3)\right|&&
    \\
    \label{e:throat-areas-ineq}
    -\left|{\bf w}(0,\zeta^2+d\zeta^2,\zeta^3)
    -{\bf w}(0,\zeta^2,\zeta^3)\right|
    \cdot
    \left|{\bf w}(0,\zeta^2,\zeta^3+d\zeta^3)
    -{\bf w}(0,\zeta^2,\zeta^3)\right|
    \;& > & 0.
\end{align}
This inequality is true for every point
${\bf w}(0,\zeta^2,\zeta^3)$. Therefore, no matter how we slice across neck 1 on either side of the 
$\zeta^1=0$ section, every nearby surface has greater area, guaranteeing that
${\bf w}(0,\zeta^2,\zeta^3)$
is a locally minimal-area throat in neck 1, satisfying Hochberg-Visser's throat definition. Since the inequality is true for every point on this throat, it also satisfies the Hochberg-Visser's strong flare-out condition.

\begin{figure}
   \begin{center}
      \includegraphics[width=1.0\linewidth]{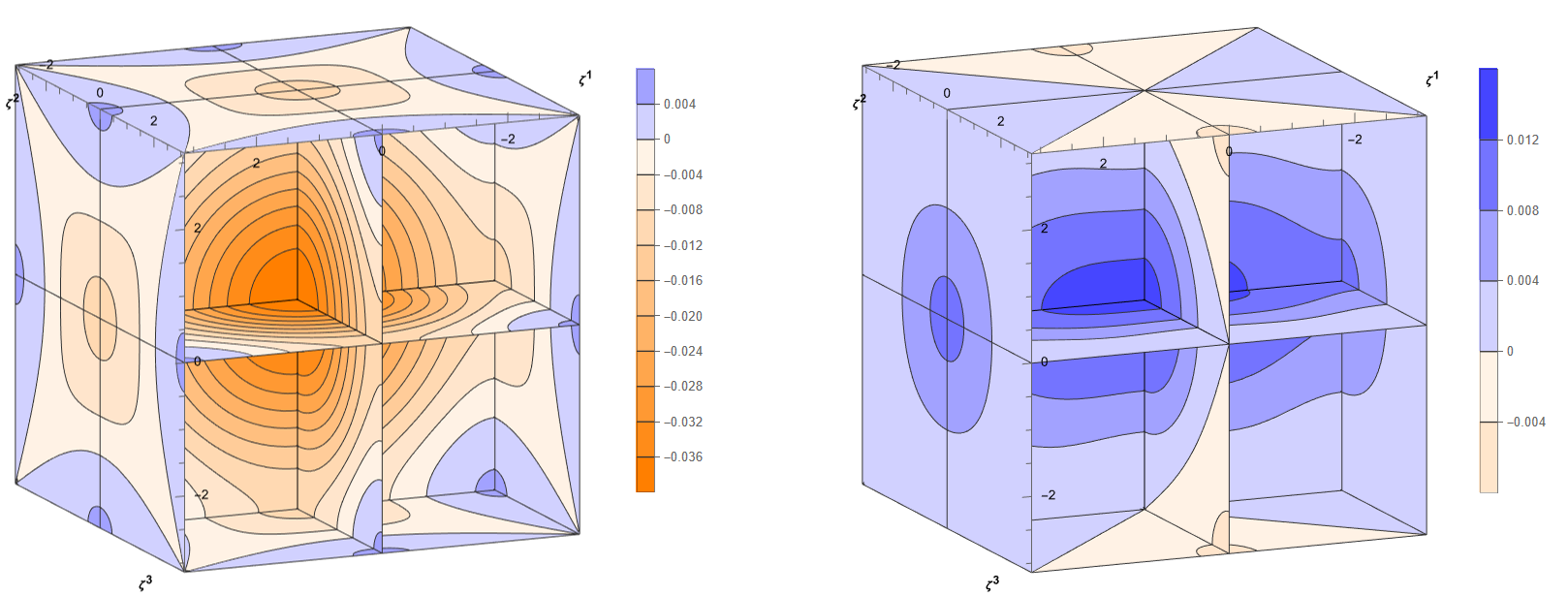}
   \end{center}
   \caption{\footnotesize
   \color{black}
   \textbf{Left: }Local inertial (tetrad) frame energy density $\hat{T}^{00}$ mapped (for visualization) back to the original coordinates domain of the flat 3-torus before inversion. Blue represents ordinary positive energy density. Orange represents poorly understood ``exotic'' negative energy density. 
   \textbf{Right: }Tetrad frame $\hat{T}^{11}$ normal stress in the $\zeta^1$ direction.
   \textbf{Both: } Scales are in units of $c^4/G\sigma^2$  using parameters $r_\mu=\sigma=1$ for $\mu=1,2,3$.
   }
   \label{f:T00}
\end{figure}

In the same manner, the section
${\bf w}(\zeta^1,0,\zeta^3)$ associated with neck 2 geodesic
${\bf w}(\pi,\zeta^2,\pi)$ and the section 
${\bf w}(\zeta^1,\zeta^2,0)$ associated with neck 3 geodesic
${\bf w}(\pi,\pi,\zeta^3)$ are locally minimal-area $2$-surfaces satisfying the same definitions.
Thus, we have three mutually orthogonal wormhole throats represented by the three orthogonal surfaces which bisect the cube in Fig. \ref{f:T00} (left) intersecting at the point ${\bf w}(0,0,0)$ represented by the cube's deep orange central point.

\noindent
$\square$

\subsection{Exotic stress-energy}\label{s:T00}

The heptad frame restricted tensors are the tetrad frame tensors of the triple wormhole 3+1-spacetime. The tetrad frame Einstein tensor is the left-hand side of the EFE (\ref{e:efe}). 
\begin{equation}\label{e:G}
  \hat{G}_{ac}
  \coloneqq
  \hat{R}_{ac}-\frac{1}{2}
  \boldsymbol{\gamma}_{ac}\hat{R}.
\end{equation}

The trace of the tetrad Ricci tensor is the tetrad Ricci scalar,
\begin{equation}\label{e:hatR2}
    \hat{R}
    =\hat{R}_a^a
    =
    -\dfrac{\frac{2}{r^1}\cos\zeta^1+\frac{2}{r^2}\cos\zeta^2+\frac{2}{r^3}\cos\zeta^3}{3\sigma\lambda}
    -\dfrac{\sin^2\zeta^1+\sin^2\zeta^2+\sin^2\zeta^3}{6\sigma^2}.
\end{equation}

In the symmetric asymptotically flat spacetime (\ref{e:w4metric}) the tetrad Ricci tensor (\ref{e:hatRii2}) and tetrad Ricii scalar (\ref{e:hatR2}) simplify to
\begin{equation}\label{e:hatRiiw}
    \hat{R}_{ii}\coloneqq 
    -\dfrac{\cos\zeta^1+\cos\zeta^2+\cos\zeta^3+\cos\zeta^i}{6\sigma^2\lambda}
    -\dfrac{\sin^2\zeta^1+\sin^2\zeta^2+\sin^2\zeta^3}{18\sigma^2}.
\end{equation}
\begin{equation}\label{e:hatRw}
    \hat{R}\coloneqq
    -\dfrac{2\cos\zeta^1+2\cos\zeta^2+2\cos\zeta^3}{3\sigma^2\lambda}
    -\dfrac{\sin^2\zeta^1+\sin^2\zeta^2+\sin^2\zeta^3}{6\sigma^2}.
\end{equation}

The tetrad Einstein tensor's nonzero components are
\begin{align}
  \label{e:hatG00}
  \hat{G}_{00}=\frac{1}{2}\hat{R}
  ,\text{ and}
\end{align}
\begin{equation}
  \label{e:hatGii}
  \hat{G}_{ii}=
    \dfrac{\cos\zeta^1+\cos\zeta^2+\cos\zeta^3-\cos\zeta^i}{6\sigma^2\lambda}
    +\dfrac{\sin^2\zeta^1+\sin^2\zeta^2+\sin^2\zeta^3}{36\sigma^2}.
\end{equation}

Since $\hat{R}^{00}=0$ (\ref{e:hatRii}), the EFE (\ref{e:efe}), (\ref{e:G}) give us the triple wormhole energy density in the local inertial frame, 
\begin{equation}\label{e:hatT00}
    \hat{T}^{00}
    =\dfrac{\hat{G}^{00}}{8\pi}
    =\dfrac{\hat{R}}{16\pi}.
\end{equation}
Setting the scaling factor $\sigma=1$ in metric (\ref{e:w4metric}), Eq.\ (\ref{e:hatT00}) produces Fig. \ref{f:T00}, a contour plot illustrating the positive and negative energy density distribution in the cube model mapped for compact illustration back onto the pre-inversion 3-torus domain. 

As in Fig. \ref{f:Ffz} (left), the eight corners of the cube model of the Fig. \ref{f:T00} (left) 3-torus are identified with the infinity point outside $\mathbb{R}^3$. Each set of four parallel cube edges are identified as the same edge. Each of the spherically inverted images of the three distinct edges, 
${\bf w}(\zeta^1,\pi,\pi)$, 
${\bf w}(\pi,\zeta^2,\pi)$, and 
${\bf w}(\pi,\pi,\zeta^3)$ are the central geodesic paths each through one the three necks. Fig. \ref{f:T00} (left) shows only positive energy density along all three paths.

The three wormhole neck throats
${\bf w}(0,\zeta^2,\zeta^3)$,
${\bf w}(\zeta^1,0,\zeta^3)$, and
${\bf w}(\zeta^1,\zeta^2,0)$
correspond to the three mutually perpendicular surfaces bisecting the 
cube in Fig. \ref{f:T00} (left) illustrating that both the greatest positive energy density and the greatest negative energy density are located on the three throats.

\subsection{Energy conditions}
\label{s:energyconditions}
The weak energy condition (WEC, \citet[pp. 721-722]{senovilla-1998}) requires that $\hat{T}_{ab}v^av^b\ge0$ for all causal vectors $\mathbf{v}$. Let $\mathbf{v}=(1,0,0,0)$. In the chosen tetrad frame $\hat{T}_{ab}=\hat{T}^{ab}$, and since $\hat{T}^{ab}$ is diagonal, it is clear that everywhere negative energy density ($\hat{T}^{00}<0$) occurs, the triple wormhole mapping (\ref{e:w4metric}) violates the WEC.

The strong energy condition (SEC, \citet[pp. 721-722]{senovilla-1998}) requires that $\hat{R}_{ab}v^av^b\ge0$ for all causal vectors $v$. Let the spatial coordinates ${\boldsymbol\zeta}=\{0,\pi,\pi\}$ which, by symmetry, is the midpoint of one neck. From (\ref{e:w}) we obtain $\lambda=3$, and from the diagonal Ricci tensor formula (\ref{e:hatRii}) we find that $\hat{R}_{00}=\hat{R}_{11}=0$ and $\hat{R}_{22}=\hat{R}_{33}=\frac{1}{9}$. Let $\mathbf{v}=(1,\delta,\delta,\delta)$, where $\delta\ne0$.
For small enough $\delta$, $\mathbf{v}$ is causal, i.e., $\gamma_{ab}v^av^b\le0$.
At coordinates ${\boldsymbol\zeta}=\{0,0,0\}$ a test particle in motion violates the SEC, but a stationary test particle ($\delta=0$) does not.

\section{Conclusions}
We have provided a global mapping equation for a multi-neck spacetime wormhole whose metric tensor is an exact solution to the Einstein field equations with a non-vanishing stress-energy tensor. Modern computers made this work less difficult than when Einstein and Rosen\cite{ER} suggested the possibility in 1935.

We proved that the triple wormhole satisfies both a global topological spacetime wormhole definition and that each neck satisfies a local geometric spacetime wormhole definition. The solution generates a diagonal stress-energy tensor. 

The asymptotically flat version of this triple wormhole solution has zero total mass, but like other spacetime wormholes in standard general relativity, the source contains regions of negative energy density, a controversial ingredient which may not physically exist. 

Misner and Wheeler's 1957 observation\cite{misner-wheeler-1957} remains true today: ``It follows that the variety of objects that can be built out of curved empty space is exceedingly rich and very far from having been explored or even surveyed.'' While the synchronous frame imposed by Eq.\ (\ref{e:g0mu}) is a simplistic choice, it is nevertheless a starting point for showing that multi-neck mathematical solutions with a stress-energy source exist. Non-synchronous-frame metrics with time dilation will follow while most of the existential questions we have about spacetime wormholes will remain difficult to answer.
 
Additional future research topics include variations of the metric tensors in this paper, 
numerical modeling, 
horizon functions,
nonstationary metrics,
analysis of deflection of light passing near or through the triple wormhole, 
and more wormhole species geometrically constructed by inverting other closed non-simply-connected manifolds.

\begin{acknowledgments}
We thank Prof. Thomas Banchoff for his groundbreaking computer graphic images of Dupin cyclides in the 1980's.
We thank Brandon Smith and David Peasley for 3-D printed models which helped in visualizing Dupin cyclides and variations of our mapping equations.

\vspace{3.5mm}
\noindent\begin{minipage}{0.15\textwidth}
\includegraphics[height=0.08\paperheight]{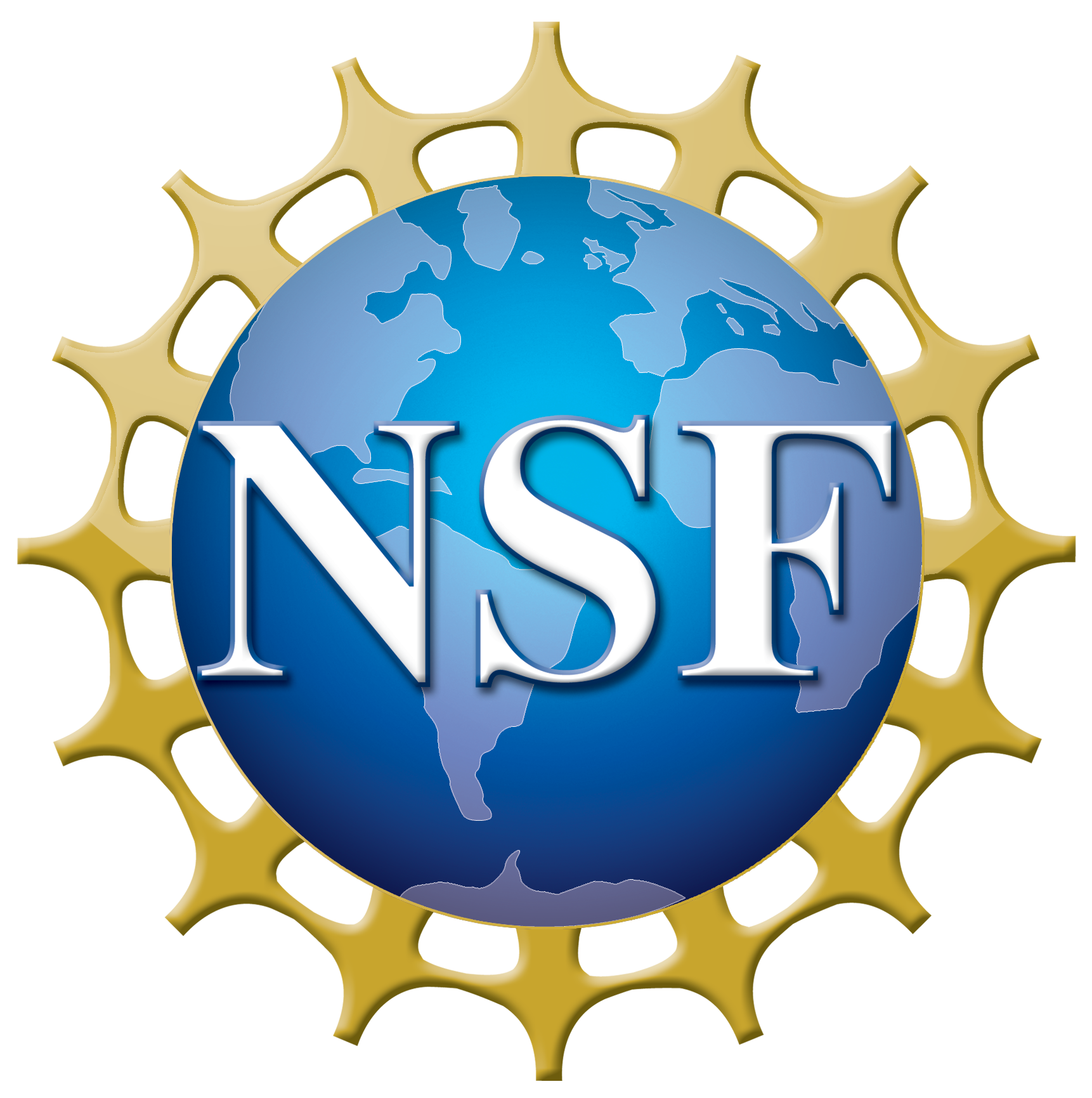}
\end{minipage}%
\hfill%
\begin{minipage}{0.8\textwidth}
This material is based upon work supported by the National Science Foundation Graduate Research Fellowship Program under Grant No. DGE-1553798. Any opinions, findings, and conclusions or recommendations expressed in this material are those of the authors and do not necessarily reflect the views of the National Science Foundation.
\end{minipage}
\end{acknowledgments}

\vspace{7mm}
\noindent
\textbf{Author Declarations}
\vspace{3.5mm}

\noindent
The authors have no conflicts to disclose.






\appendix
\section{Coordinate frame geometric formulas}
\label{s:APXCgeom}

\subsection{Einstein field equations (EFE)}
\label{s:APXefe}

%
The Einstein field equations (\ref{e:APXefe}) are a system of PDEs that equate the Einstein tensor $G_{\mu\nu}$ with the density, stresses, and momentum of matter-energy sources expressed in the stress-energy tensor $T_{\mu\nu}$ (times a universal constant):
\begin{equation}\label{e:APXefe}
    G_{\mu\nu}
    \coloneqq R_{\mu\nu}-\frac{1}{2}g_{\mu\nu}R
    =8\pi 
    T_{\mu\nu} 
\end{equation}
(\citet[Eq.\ (7.1.13)]{weinberg-1972} using the positive Riemann sign convention\cite{mtw} and geometrized units $c=1$ and $G=1$).
Einstein replaced classical gravitational fields by spacetime curvature expressed in terms of the coordinate frame metric tensor $g_{\mu\nu}
$, the coordinate frame Ricci curvature tensor
(\citet[Eq.\ (6.2.4)]{weinberg-1972})
\begin{equation}\label{e:APXricci}
    R_{\mu\nu}
    \coloneqq{R^\lambda}_{\mu\lambda\nu},
\end{equation}
and the coordinate frame curvature scalar (\citet[Eq.\ (6.2.5)]{weinberg-1972})
\begin{equation}\label{e:APXR}
    R
    \coloneqq g^{\mu\nu}R_{\mu\nu},
\end{equation}
where the coordinate frame Riemann curvature tensor
(\citet[Eq.\ (6.1.5)]{weinberg-1972} using the MTW sign conventions\cite{mtw}) is
\begin{equation}\label{e:APXriemann}
    {R^\lambda}_{\mu\nu\kappa}
    \coloneqq
    \partial_\nu\varGamma^\lambda_{\mu\kappa}
    -\partial_\kappa\varGamma^\lambda_{\mu\nu}
    +\varGamma^\eta_{\mu\kappa}\varGamma^\lambda_{\nu\eta}
    -\varGamma^\eta_{\mu\nu}\varGamma^\lambda_{\kappa\eta}.
\end{equation}
The Christoffel symbol (\citet[Eq.\ (3.3.7)]{weinberg-1972}) is
\begin{equation}\label{e:APXChristoffel}
    \varGamma^\kappa_{\lambda\mu}
    \coloneqq
    \frac{1}{2}g^{\nu\kappa}
    \left(\partial_\lambda g_{\mu\nu}
    +\partial_\mu g_{\lambda\nu}
    -\partial_\nu g_{\mu\lambda}\right),
\end{equation}
and the inverse metric $g^{\nu\kappa}
$ is defined everywhere by (\citet[Eq.\ (3.3.6)]{weinberg-1972})
\begin{equation}\label{e:APXinvg}
    g^{\nu\sigma}g_{\mu\nu}=\delta^\sigma_\mu.
\end{equation}

\subsection{Parabolic Dupin cyclides in the triple wormhole}
\label{s:APXPDC}
A rigid rotation of the coordinate surface 
$\zeta^3=\pi$ in mapping (\ref{e:w})

\begin{equation}\label{e:APXrot}
    \begin{bmatrix}
        0 & 1 & 0 & 0 && 0 && 0 \\
        0 & 0 & 0 & 1 && 0 && 0 \\
        -\frac{1}{\sqrt{2}} & 0 & \frac{1}{\sqrt{2}} & 0 && 0 && 0 \\
        -\frac{1}{\sqrt{2}} & 0 & -\frac{1}{\sqrt{2}} & 0 && 0 && 0 \\
        0 & 0 & 0 & 0 && 1 && 0 \\
        0 & 0 & 0 & 0 && 0 && 1
    \end{bmatrix}
    \mathbf{w}(\zeta^1,\zeta^2,\pi)
    =
    6\sigma
    \begin{bmatrix}
        \frac{\sin\zeta^1}
        {2+\cos\zeta^1+\cos\zeta^2} \\
        \frac{\sin\zeta^2}
        {2+\cos\zeta^1+\cos\zeta^2} \\
        \frac{\cos\zeta^2-\cos\zeta^1}
        {\sqrt{2}(2+\cos\zeta^1+\cos\zeta^2)} \\
        -\frac{1}{\sqrt{2}} \\
        0 \\
        0
    \end{bmatrix},
\end{equation}
and plotting the three non-constant coordinates shows that this 2-manifold is a parabolic Dupin cyclide  (Figs. \ref{f:dupin} and \ref{f:geodesics}) that can be isometrically embedded in $\mathbb{R}^3$. Likewise for the other two coordinate surfaces $\zeta^\alpha=\pi$.

While the following transformation (\ref{e:APXrot2}) is not a rigid rotation, it indicates that certain diagonal sections of mapping (\ref{e:w}) intersecting the origin are asymptotically flat 2-surfaces with the same topology as the PDC, and with planar cross-sections being elliptical instead of circular:
\begin{equation}\label{e:APXrot2}
\begin{bmatrix}
0 & 1 & 0 & 0 & 0 & 0 \\
 0 & 0 & 0 & \frac{1}{\sqrt{2}} & 0 & \frac{1}{\sqrt{2}} \\
 \frac{2}{\sqrt{5}} & 0 & -\frac{1}{\sqrt{5}} & 0 & 0 & 0 \\
 \frac{1}{\sqrt{5}} & 0 & \frac{2}{\sqrt{5}} & 0 & 0 & 0 \\
 0 & 0 & 0 & \frac{1}{\sqrt{2}} & 0 & -\frac{1}{\sqrt{2}} \\
 0 & 0 & \frac{1}{\sqrt{2}} & 0 & -\frac{1}{\sqrt{2}} & 0
\end{bmatrix}
\mathbf{w}(\zeta^1,\zeta^2,\zeta^2)
=
6\sigma
\begin{bmatrix}
 \frac{\sin\zeta^1}{3+\cos\zeta^1+2\cos\zeta^2} \\
 \frac{\sqrt{2}\sin\zeta^2}{3+\cos\zeta^1+2\cos\zeta^2} \\
 \frac{1+2\cos\zeta^1-\cos\zeta^2}{\sqrt{5}\left(3+\cos\zeta^1+2\cos\zeta^2\right)}\\
 \frac{1}{\sqrt{5}} \\
 0 \\
 0
\end{bmatrix}.
\end{equation}

\section{Tetrad frame geometric formulas}
\label{s:APXTgeom}
%
\color{black}
\subsection{Spatial hexads}
\label{s:APXhexad}
Pre-inversion coordinate system (\ref{e:f}) foliates $\mathbb{R}^6$ with embedded flat 3-torus manifolds. We chose simple sparse orthonormal hexad vectors $\boldsymbol{\alpha}_m$ for every point in $\mathbb{R}^6\setminus0$, where 
$\boldsymbol{\alpha}_m=\partial{\bf f}/\partial x^m/\sqrt{}((\partial{\bf f}/\partial x^m)\cdot(\partial{\bf f}/\partial x^m))$ is just the usual tangent basis of (\ref{e:f}). The first three hexad vectors are tangent to the embedded 3-torus manifold passing through each point. The next three are normal to the manifold.
The specific values we used for a hexad of orthonormal local basis vectors on a flat 3-torus embedded in $\mathbb{R}^6$ are
  \scriptsize
\begin{equation}\label{e:APXfhexad}
  \boldsymbol{\alpha}_1
  =\begin{bmatrix}
    -\sin\zeta^1\\ \cos\zeta^1\\0\\0\\0\\0
  \end{bmatrix},
  \boldsymbol{\alpha}_2
  =\begin{bmatrix},
    0\\0\\-\sin\zeta^2\\\cos\zeta^2\\0\\0
  \end{bmatrix},
  \boldsymbol{\alpha}_3
  =\begin{bmatrix}
    0\\0\\0\\0\\ -\sin\zeta^3\\ \cos\zeta^3
  \end{bmatrix},
  \boldsymbol{\alpha}_4
  =\begin{bmatrix}
    \cos\zeta^1\\ \sin\zeta^1\\0\\0\\0\\0
  \end{bmatrix},
  \boldsymbol{\alpha}_5
  =\begin{bmatrix}
    0\\0\\ \cos\zeta^2\\ \sin\zeta^2\\0\\0
  \end{bmatrix},
  \boldsymbol{\alpha}_6
  =\begin{bmatrix}
    0\\0\\0\\0\\ \cos\zeta^3\\ \sin\zeta^3
  \end{bmatrix}.
\end{equation}
  \normalsize
where $\zeta^\mu$ are the flat 3-torus coordinates.

The post-inversion hexad vectors $\boldsymbol{\beta}_m$ are reflections of the corresponding pre-inversion hexad vectors $\boldsymbol{\alpha}_m$ across a hyperplane tangent to the inversion sphere
at the point where it is intersected by a ray from the inversion sphere's center ($\mathbb{R}^6$ origin) through the point ${\bf f}$ corresponding by relationship (\ref{e:z}) to the point ${\bf z}$.

The $6\times6$ Householder matrix $\mathsf{H}$ describing this transformation is calculated by
\begin{equation}\label{e:APXH}
    \mathsf{H}
    =\mathsf{I}
    -\dfrac{2{\bf f}{\bf f}^\intercal}
    {\left|{\bf f}\right|^2},\quad
    \text{to give us the reflection: }
    \mathsf{H}{\bf f}=-{\bf f}.
\end{equation}
\begin{equation}\label{e:APXzhexad}
    \boldsymbol{\beta}_m
    =\mathsf{H}\boldsymbol{\alpha}_m.
\end{equation}
$\boldsymbol{\beta}_m=\partial{\bf z}/\partial x^m/\sqrt{}((\partial{\bf z}/\partial x^m)\cdot(\partial{\bf z}/\partial x^m))$ is the orthonormal tangent basis of the inverted mapping (\ref{e:z}). 
From (\ref{e:APXzhexad}) the specific hexad vector values that we assigned to (\ref{e:z}) are
\begin{align}
  \nonumber
  \boldsymbol{\beta}_1 
  &= 
  \boldsymbol{\alpha}_1
  + \dfrac{\sin\zeta^1}
  {6\sigma}
  \lambda{\bf f},
  &\boldsymbol{\beta}_4 
  &= 
  \boldsymbol{\alpha}_4
  -\dfrac{\frac{r^1}{\sigma}+\cos\zeta^1}
  {6\sigma}
  \lambda{\bf f},
  \\
  \nonumber
  \boldsymbol{\beta}_2 
  &= 
  \boldsymbol{\alpha}_2
  +\dfrac{\sin\zeta^2}
  {6\sigma}
  \lambda{\bf f},
  &\boldsymbol{\beta}_5 
  &= 
  \boldsymbol{\alpha}_5
  -\dfrac{\frac{r^2}{\sigma}+\cos\zeta^2}
  {6\sigma}
  \lambda{\bf f},
  \\
  \label{e:APXzhexad2} 
  \boldsymbol{\beta}_3 
  &= 
  \boldsymbol{\alpha}_3
  +\dfrac{\sin\zeta^3}
  {6\sigma}
  \lambda{\bf f},
  &\boldsymbol{\beta}_6 
  &= 
  \boldsymbol{\alpha}_6
  -\dfrac{\frac{r^3}{\sigma}+\cos\zeta^3}
  {6\sigma}
  \lambda{\bf f},
\end{align}
where $\sigma$ is a scaling factor and
$\lambda$ is the conformal factor (\ref{e:lambda-z}).
 
\subsection{Spacetime heptads}
\label{s:APXheptadmetric}
When orthonormal heptads are chosen, as above, the heptad metric equals the Minkowski metric, $\gamma_{mn}=\eta_{mn}$. In this paper, heptad indices are Latin letters where the early letters $a,b,c,\dots$ can take values only in $\{0,1,2,3\}$, late letters $w,x,y,z$ can take values in $\{4,5,6\}$, and middle letters $k,l,m,n$ can take any integer value in $\{0,1,\cdots,6\}$. We reserve letters $i$ and $j$ for nonzero values in $\{1,2,3\}$.

We define the triple wormhole's heptad vectors by
\begin{equation}
    \gamma_0=\left(1,0,0,0,0,0,0\right),\quad
    \gamma_m=\beta_m,\quad
    m\in\{1,2,3,4,5,6\}.
\end{equation}
In general, the \emph{heptad metric} is the heptad vectors' dot product (\citet[Eq.\ (11.3)]{hamilton-grbook}):
\begin{equation}\label{e:APXheptadmetric}
   \gamma_{mn}
   \coloneqq
   \boldsymbol{\gamma}_m\cdot\boldsymbol{\gamma}_n
   \coloneqq
   (\boldsymbol{\gamma}_m)^\intercal
   \eta\,\boldsymbol{\gamma}_n
\end{equation}
where the dot product is defined by the (Minkowski) norm of the ambient spacetime.

The heptad inverse metric is defined everywhere by 
$\gamma_{km}\gamma^{mn}=\delta_k^n$ \citet[Eq.\ (11.19)]{hamilton-grbook}.

\subsection{Tetrad-frame connection}

\label{s:inv-sieb-deriv}
Specific values for the inverse siebenbein derivative (\ref{e:dlmn}), not a tensor, are
\begin{align}
    \nonumber
    d_{llz} 
    =& 
    -\dfrac{r^i}{6\sigma^2}
    -\dfrac{\cos\zeta^i}
    {6\sigma}
    +\left\lbrace
    \begin{array}{ll}
      \quad 0,
      &  \quad
      i=z-3\ne l\ne0
      \\
      \dfrac{1}{\lambda r^i},
      & \quad
      i=z-3=l\ne0,
    \end{array}
    \right.
    \\
    \label{e:APXdlmn1}
    d_{llj} 
    =&
    \dfrac{\sin\zeta^j}
    {6\sigma},\quad
    j\ne0,
    l\ne0;\quad 
    \text{ all other }\;
    d_{lmn} = 0.
\end{align}

\subsection{Tetrad tensors}
\label{s:APXtetradtensors}

In general, the polyad-frame \emph{Riemann tensor} is calculated by \citet[Eq.\ (11.61)]{hamilton-grbook}:
\begin{equation}\label{e:APXRklmn}
    R_{klmn}\coloneqq
    \partial_k\Gamma_{mnl}
    -\partial_l\Gamma_{mnk}
    +\Gamma^p_{ml}\Gamma_{pnk}
    -\Gamma^p_{mk}\Gamma_{pnl}
    +(\Gamma^p_{kl}-\Gamma^p_{lk})\Gamma_{mnp}=0
    ,
\end{equation}
where the heptad connections $\Gamma$ are calculated by (\ref{e:heptadconnection}) with indices raised by the heptad inverse metric. For the case described in this paper, this tensor (\ref{e:APXRklmn}) vanishes everywhere, see (\ref{e:gauss}). 

The Riemann tensor (\ref{e:APXRklmn}), is restricted to the tangent tetrad spacetime indices $0, 1, 2,$ and $3$ using \citet[Eq.\ (17.35)]{hamilton-grbook}.
\begin{equation}\label{e:APXhatRabcd}
    \hat{R}_{abcd}\coloneqq
    \partial_a\hat{\Gamma}_{cdb}
    -\partial_b\hat{\Gamma}_{cda}
    +\hat{\Gamma}^p_{cb}\hat{\Gamma}_{pda}
    -\hat{\Gamma}^p_{ca}\hat{\Gamma}_{pdb}
    +(\Gamma^p_{ab}-\Gamma^p_{ba})\hat{\Gamma}_{cdp}
    .
\end{equation}


\bibliography{TripleWormhole}

\end{document}